\documentclass[12pt]{article}
\pdfoutput=1
\usepackage{amssymb,amsmath,empheq,epsfig,fixltx2e,hyperref,psfrag,simplewick}
\graphicspath{{./figures/}}
 
\numberwithin{equation}{section}

\newcommand{\be}{\begin{equation}}
\newcommand{\bea}{\begin{eqnarray}}
\newcommand{\ee}{\end{equation}}
\newcommand{\eea}{\end{eqnarray}}
\newcommand{\ret}{\nonumber \\}
\newcommand{\nn}{\nonumber}
\newcommand{\refb}[1]{(\ref{#1})}

\newcommand{\cO}{\mathcal{O}} 
\newcommand{\cN}{\mathcal{N}} 
\newcommand{\cR}{\mathcal{R}} 

\newcommand{\mR}{\mathbb{R}} 

\newcommand{\ml}[1]{\begin{multline} #1 \end{multline}}

\newcommand{\bra}[1]{\langle#1|}
\newcommand{\ket}[1]{|#1\rangle}
\newcommand{\braket}[2]{\langle#1|#2\rangle}
\newcommand{\cor}[1]{\langle#1 \rangle}
\newcommand{\corb}[1]{\langle\langle#1 \rangle\rangle}
\newcommand{\norm}[1]{\parallel#1\parallel}
\def\tr{ {\rm tr} }
\def\str{ {\rm Str} }
\newcommand{\hot}[2]{ \cO\left(\frac{#1}{#2}\right)}

\newcommand{\stirling}[2]{\genfrac{[}{]}{0pt}{}{#1}{#2}}

\newcommand{\boxeq}[1]{\begin{empheq}[box={\fboxsep=10pt\fbox}]{align} #1 \end{empheq}}

\begin{document}

\rightline{QMUL-PH-13-15}

\vspace*{2cm} 

{\LARGE{  
\centerline{   \bf Thresholds of Large $N$ Factorization in CFT\textsubscript{4}: } 
\centerline{ \bf Exploring bulk spacetime in AdS\textsubscript{5} } 
}}

\vskip.5cm 

\thispagestyle{empty} \centerline{
    {\large \bf David Garner,
${}^{a,} $\footnote{ {\tt d.p.r.garner@qmul.ac.uk}}}
   {\large \bf Sanjaye Ramgoolam,
               ${}^{a,}$\footnote{ {\tt s.ramgoolam@qmul.ac.uk}}   }
   {\large \bf and Congkao Wen
               ${}^{a,}$\footnote{ {\tt c.wen@qmul.ac.uk}}   }
                                                       }

\vspace{.4cm}
\centerline{{\it ${}^a$ Centre for Research in String Theory,}}
\centerline{ {\it School of Physics and Astronomy},}
\centerline{{ \it Queen Mary University of London},} 
\centerline{{\it    Mile End Road, London E1 4NS, UK}}

\vspace{1.4truecm}

\thispagestyle{empty}

\centerline{\bf ABSTRACT}
\vspace*{1cm} 

Large $N$ factorization ensures that, for low-dimension gauge-invariant operators in the half-BPS sector of $\cN=4$ SYM,
products of holomorphic traces have vanishing correlators with single anti-holomorphic traces. 
This vanishing is necessary to consistently map trace operators in the CFT$_4$ to a Fock space of graviton oscillations 
in the dual AdS$_5$.
We investigate the regimes at which the CFT correlators do not vanish but become of order one in the large $N$ limit, which we call a {\it factorization threshold}.
Quite generally, we find the threshold to be when the product of the two holomorphic operator dimensions is of order $N\log N$. 
Our analysis considers extremal and non-extremal correlators and correlators in states dual to LLM backgrounds, and we 
observe intriguing similarities between the the energy-dependent running coupling of non-abelian gauge theories and our 
threshold equations.
Finally, we discuss some interpretations of the threshold within the bulk AdS spacetime.

\vskip.4cm

\setcounter{page}{0}
\setcounter{tocdepth}{2}

\newpage 

\tableofcontents

\section{Introduction}

In the AdS/CFT correspondence \cite{malda,GKP,witten},
chiral primary operators of small dimension in $\cN=4$ super Yang-Mills theory are
dual to Kaluza-Klein graviton excitations in Type IIB supergravity on $AdS_5\times S^5$ \cite{witten}.
A remarkable early success of AdS/CFT was the explicit large $N$ calculation and matching of
the three-point correlators of gauge theory operators with the associated graviton correlators in supergravity \cite{LMRS}. On the gauge theory side, the  operators are symmetric traceless combinations of 
the six adjoint scalar fields, and the correlator can be calculated 
at zero gauge coupling $g_{YM}^2$.
On the supergravity side, the corresponding fields arise from the Kaluza-Klein
reduction along the 5-sphere of excitations of the metric and the self-dual 5-form field strength.
The agreement between these correlators on both sides of the correspondence is possible because three-point
functions of chiral primary operators are not renormalized \cite{NonRen1,NonRen2,NonRen3,NonRen4,NonRen5,NonRen6}.

The half-BPS sector of chiral primary operators is described by a single holomorphic matrix 
$Z=\Phi_1+i\Phi_2$, formed from the complex combination of two adjoint hermitian scalars \cite{bbns, cjr, berenstein}.
Single trace operators consisting of a small number of $Z$ matrices can be matched to single particle bulk graviton states,
and multi-trace operators can be matched to multi-graviton states.
The number of matrices $J$ in a single trace operator corresponds to the angular momentum of the Kaluza-Klein graviton
in the $S^5$ directions.
For a three-point extremal correlator of the form 
\bea
\cor{\tr Z^{J_1}(x_1) \tr Z^{J_2}(x_2) \tr Z^{\dagger J_1+J_2}(y)},
\eea
the conformal symmetry allows the spacetime dependence of the correlator to be factored out completely.
The remaining factor is purely combinatoric, and gives the CFT inner product between the 2-graviton state
and the 1-graviton state. This combinatoric factor is known exactly for finite $N$, arising directly from Wick contraction 
combinatorics of matrices \cite{kpss,cr}.  
With an appropriate normalization, this free-field correlator goes to zero 
in the limit of large $N$ when the operator dimensions $J_i$ ($i=1,2$) are kept fixed.  This is an example 
of a general property of large $N$ physics called {\it large $N$ factorization}. 
It is necessary for a weakly-coupled Fock space description of the bulk theory to be valid. 
The single trace operators can be matched with a set of graviton oscillators 
\bea 
tr Z^J \leftrightarrow \alpha_{J},
\eea
with the commutation relations $[ \alpha_{ J_1 }, \alpha_{ J_2 }^{ \dagger} ] = \delta_{ J_1, J_2 }$, 
which annihilate the AdS vacuum state $\alpha_J\ket{0}=0$. The excitations of the vacuum state
form a Fock space, and correlators of states with different numbers of excitations are orthogonal:
\bea 
\bra{0} \alpha_{J_1} \alpha_{J_2} \alpha_{J_1 + J_2}^{ \dagger} \ket{0}  = 0,
\eea
which is in agreement with the CFT correlator at large $N$.

In this paper we will be interested in the growth of the operator dimensions $J_i$ which 
leads to the failure of factorization.  We find that if the $J_i$ grow sufficiently 
rapidly with $N$, the normalized correlator diverges as $N \rightarrow \infty$. 
We undertake a detailed study of the {\it factorization threshold}, defined to be the
submanifold of the space of parameters (dimensions and global symmetry charges of the operators,  and $N$) on which the normalized correlator is equal to a constant $c$, chosen for convenience to be $1$ in most formulae.  
At the threshold, $c$ can be as small as we like but independent of $N$.  
It therefore makes sense in this regime to associate single traces to single objects and multi-traces to multiple objects, just  as it does below the threshold. 
However, a  standard Fock space structure as the starting point for a $1/N$ expansion 
is not appropriate at the  threshold. 
Here, composite states made of a pair of gravitons have non-vanishing quantum correlations with states consisting of a single graviton, even as $N$ is taken to infinity. 
This motivates the detailed characterization and interpretation of the threshold, which we undertake in this paper. 
Above the threshold, associating single traces to single objects of any sort probably does not make sense.
Certainly, for $J_i$ of order $N$, it is known that the gravitons are represented semiclassically by D3-branes wrapping a sphere \cite{mst},
and cannot be represented as single traces \cite{bbns}. The correct basis for single and giant gravitons 
is given by Schur polynomials, indexed by Young diagrams \cite{cjr}.

The aim of this paper is to introduce and investigate the threshold of factorization
for several cases of correlators in the half-BPS sector of $\cN=4$ super Yang-Mills,
and to explore the implications  of these in the dual $AdS_5\times S^5$ spacetime. 
We focus on three types of correlator in particular: an extremal three-point correlator with 
one independent angular momentum $J$, an extremal three-point correlator with 
two independent angular momenta $J_1$ and $J_2$, and a non-extremal three-point 
correlator. We also consider briefly some extensions concerning extremal correlators on 
non-trivial backgrounds and extremal correlators with a large number of operators.

In Section \ref{sec:overview} we give an overview of our results, introducing
the definition of the factorization threshold and stating without detailed calculation 
the form of the threshold in the simplest case.  The local gauge invariant operators 
are functions of a four-dimensional spacetime position and an energy $J$, which is equal to 
angular momentum because of the BPS condition. We explain an interesting aspect of 
our results, namely the similarity of the dependence of the threshold on separations 
in spacetime and on differences in energy.  We elaborate on the departure from the 
usual Fock space structure associated with traces at large N and raise the question of 
a spacetime effective field theory derivation of the properties of the threshold. 
This is one of our motivations for performing detailed studies of the threshold.

In the subsequent sections, we present the details of the calculations of the thresholds.
In Section \ref{sec:j1eqj2} we review and introduce some notation 
on large $N$ asymptotics for describing the thresholds precisely, 
and give a complete calculation of the extremal three-point correlator
with one independent angular momentum $J$. We also discuss in this section some links between 
the form of the threshold equations with running gauge coupling equations and instanton expansions.
In Section \ref{sec:j1neqj2}, we present a calculation of the three-point extremal correlator
when the operator dimensions are not equal. 
In Section \ref{sec:next}, we calculate a non-extremal three-point correlator,
and discuss how it differs from the extremal cases.

We discuss in Section \ref{sec:future} some other tractable examples of extremal correlators that could shed more 
light on the general nature of factorization thresholds.
We consider the case of a correlator with $k$ holomorphic insertions, 
and also the case of a three-point correlator on a non-trivial background dual to an LLM geometry \cite{LLM}.
We conclude by summarizing what has been shown about factorization within this paper,
and discussing some other examples of correlators that could tell us more about the general nature of factorization 
thresholds in the future.

\section{Factorization thresholds and bulk interpretations} \label{sec:overview}

In this section we describe the factorization threshold for the simplest case: the transition
of two gravitons with the same angular momentum $J$ going to a single graviton of
angular momentum $2J$. 
This is followed by a discussion of the physics at the threshold in the bulk AdS space.
This motivates further investigations of thresholds, which we outline, along with the qualitative results. 
The details of these investigations are presented in subsequent sections.

\subsection{Thresholds of factorization in the gauge theory}

Our starting point is the three-point correlator of two holomorphic single trace operators and an antiholomorphic single trace operator,
\bea
\cor{\tr Z^{J_1}(x_1)\tr Z^{J_2}(x_2)\tr Z^{\dagger J_1+J_2}(y)}.
\eea
This correlator is not renormalized \cite{LMRS}, and so a calculation in the free field limit will hold for all values of 
the coupling $g_{YM}^2$. 
The position-dependence of the correlator can be factored out by conformal symmetry:
\bea
\cor{\tr Z^{J_1}(x_1)\tr Z^{J_2}(x_2)\tr Z^{\dagger J_1+J_2}(y)} = \frac{\cor{\tr Z^{J_1} \tr Z^{J_2} \tr Z^{\dagger J_1+J_2} }}{|x_1-y|^{2J_2}|x_2-y|^{2J_2}}.
\eea
The factor in the numerator of this expression is position-independent and can be calculated using character expansions \cite{kpss}.
If we apply an inversion  $y^{\prime } = \frac{y}{|y|^2 }$, and transform  the anti-holomorphic operator to 
the primed frame, while taking $x \rightarrow 0, y^{\prime} \rightarrow 0$, then the position dependence disappears, 
and we are left with the purely combinatoric factor which can be interpreted as an inner product of the double trace state 
and the single trace state. This correlator is \emph{extremal} as the sum of the holomorphic operator dimensions $J_1+J_2$ is equal to the antiholomorphic operator dimension. In the following sections, we focus  on the inner product 
\bea
\cor{\tr Z^{J_1} \tr Z^{J_2} \tr Z^{\dagger J_1+J_2} }.
\eea
A natural normalization for these correlators is the \emph{multiparticle normalization}, in which each operator is divided by the square root of its two-point function,
\bea
\corb{\tr Z^{J_1}\tr Z^{J_2} \tr Z^{\dagger J_1+J_2}} &=&\frac{ \cor{\tr Z^{J_1}\tr Z^{J_2}\tr Z^{\dagger J_1+J_2}} }{\sqrt{\cor{\tr Z^{J_1}\tr Z^{\dagger J_1}} \cor{\tr Z^{J_2}\tr Z^{\dagger J_2}} \cor{\tr Z^{J_1+J_2}\tr Z^{\dagger J_1+J_2}} }}.
\eea 
This normalization is used in comparing AdS and CFT calculations of the 3-point functions \cite{LMRS}.
We have introduced the double-bracket notation $\corb{\cdot}$ to refer to a multiparticle-normalized correlator.
It is known \cite{LMRS,dms} that when the operator dimensions $J_i$ are sufficiently small, then
\bea
\corb{\tr Z^{J_1}\tr Z^{J_2} \tr Z^{\dagger J_1+J_2}} \sim \frac{\sqrt{J_1J_2(J_1+J_2)}}{N} \label{eq:orthog}
\eea
in the large $N$ limit. This clearly tends to zero at large $N$, and so the single trace and double trace operators
are orthogonal at large $N$. 

Large $N$ orthogonality of the operators can still hold when $J_1$ and $J_2$ increase with $N$. By calculating the
correlator explicitly at finite $N$, it can be shown that \refb{eq:orthog} is still valid when
$J_1$ and $J_2$ are functions of $N$, provided that $J_1, J_2 \leq \sqrt{N}$ at large $N$.
However, this formula is not valid when $J_1$ and $J_2$ grow large enough with $N$. 
For large enough $J_i$, the normalized correlator grows exponentially with $N$, and factorization of the operators
no longer holds. The aim of this paper is to investigate and interpret the threshold partitioning these two 
distinct large $N$ limits of the normalized correlator.

For simplicitly, we initially consider in Section \ref{sec:j1eqj2} a correlator in which the holomorphic operator dimensions are equal.
Setting $J_1=J_2=J$, we define 
\bea
G_3(J, N) = \corb{\tr Z^J \tr Z^J \tr Z^{\dagger 2J}}.
\eea
To gain some insight into the large $N$ behaviour of this correlator when $J$ depends on $N$, we can 
plug in a simple trial function $J(N)$ and find the asymptotic behaviour of the correlator when $N$ is large.
If we set $J=N^\alpha$, where $\alpha$ is a constant, then a finite $N$ calculation \cite{dms} shows that
\boxeq{
G_3(N^\alpha, N) \to 0, &\qquad 0<\alpha\leq \frac{1}{2}, \ret
G_3(N^\alpha, N) \to \infty, &\qquad \frac{1}{2}<\alpha<1. \label{eq:box1}
}
If $J$ grows as a power of $N$ larger than $\frac{1}{2}$, then the correlator will diverge and factorization breaks down. 
However, a simple power-law scaling is not sufficient to deduce the exact growth of $J$ that is required for the correlator to diverge.
A more general $N$-dependence can be found, intermediate between the cases $\alpha=\frac{1}{2}$ and $\alpha>\frac{1}{2}$,
for which the correlator tends to a constant value. 

\begin{figure}[t]
\centering
\includegraphics[width=0.5\textwidth]{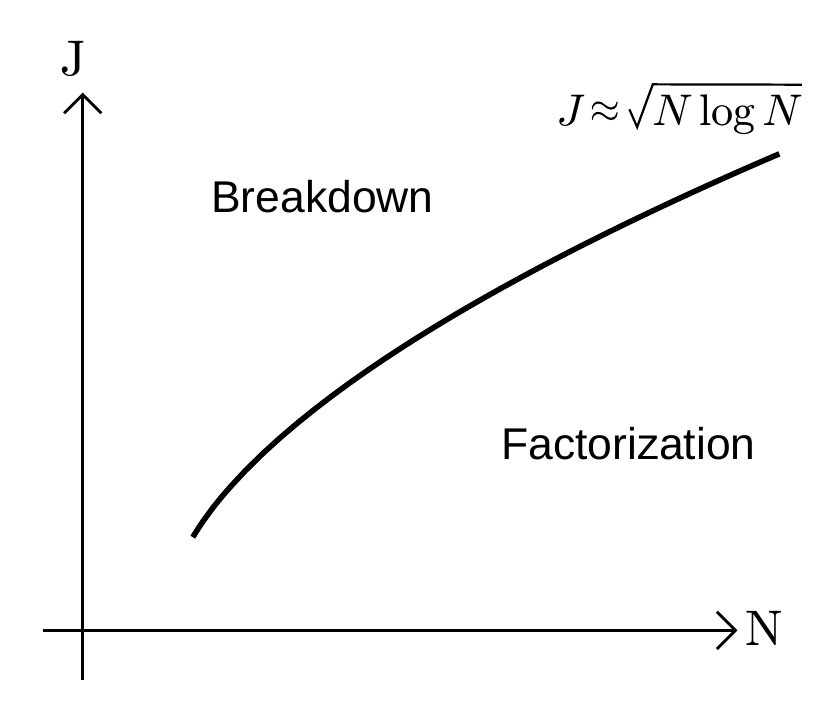}
\caption{A sketch of the threshold curve $J_t(N)$ in $(J, N)$ parameter space for large $N$. Away from the origin, the
curve partitions the parameter space into the factorization region $G_3(J_t(N), N)<1$, and the breakdown region 
$G_3(J_t(N), N) > 1$.}
\label{fig:factbreakdown}
\end{figure}

Our main approach to considering the threshold between factorization and breakdown is to look for a solution to
the equation
\bea
G_3(J, N) = 1. 
\eea
We call this the \emph{factorization threshold equation}. It defines a curve
$J(N)$ in the parameter space with axes labelled $(J, N)$. For large enough $N$, this curve 
divides the parameter space into two regions: the factorization region, where the correlator is less than one, 
and the breakdown region, where the correlator is greater than one.
The \emph{threshold} $J_t(N)$ is the exact solution of the equation $G_3(J_t(N), N) = 1$.
A sketch of this threshold curve in $(J,N)$ parameter space is shown in Figure \ref{fig:factbreakdown}.

The trial function approach shows that the threshold must scale with $N$ at a faster rate than $\sqrt{N}$, but at
a slower rate than $N^{\frac{1}{2}+\delta}$ for any constant $\delta$.
Provided that $J$ lies in the range $N^\frac{1}{2} < J < N^{\frac{1}{2}+\delta}$, we show in Section \ref{sec:asymptotics} that the correlator $G_3$ has the asymptotic behaviour
\bea
G_3(J,N)\sim  \sqrt{\frac{J}{2N}}\exp\left(\frac{J^2}{2N}\right).
\eea
Using this asymptotic form of the correlator, we can invert the equation $G_3(J_t(N), N) =1$ 
to derive an asymptotic solution of $J_t(N)$, the threshold of factorization. In Section \ref{sec:j1eqj2soln} we show that
the large $N$ solution is 
\bea
J_t(N) = \sqrt{\frac{1}{2}N\log N}\left[1 - \frac{\log \log N}{2\log N} + \frac{\log 8}{2\log N} + \hot{\log \log N}{\log N}^2 \right].
\eea
Neglecting the constant term, the leading-order behaviour is simply
\bea
J_t^2 \approx N\log N.
\eea
This is the solution that divides $(J,N)$-space into the regions where factorization holds and breaks down.

\subsection{The breakdown of bulk effective field theory at the threshold}

\begin{figure}[t]
\centering
\includegraphics[width=0.65\textwidth]{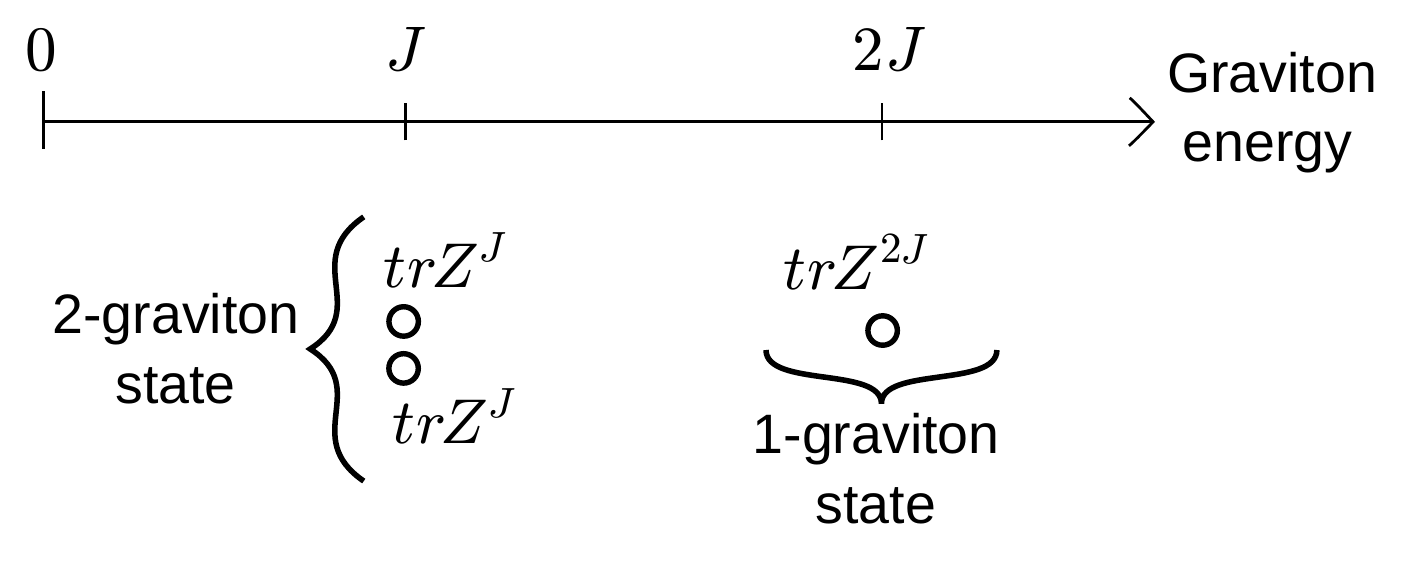}
\caption{The single and multi-graviton states within energy space.}
\label{fig:bulkbdy}
\end{figure}

The correlator $\cor{\tr Z^J \tr Z^J  ( \tr Z^{\dagger} )^{2J}}$ is not renormalized \cite{LMRS}. 
It is an inner product of the double trace state with  the single trace state, normalized by the 
appropriate factors given above. A sketch of these two states in energy space is given in Figure \ref{fig:bulkbdy}.
In the CFT computation, this is a non-trivial inner product 
which mixes trace structures according to a non-trivial function of $J$ and $N$. This inner product 
can equally be computed for $J$ of order one in the large $N$ 
limit in the dual supergravity. The supergravity computation can be understood as relying 
on a Fock space structure for gravitons, where at leading large $N$ single gravitons 
are orthogonal to multi-gravitons, hence single traces are orthogonal to multi-traces. 
This Fock space structure is used to set up perturbation theory where there are $\frac{1}{N}$ 
interactions. The $N$-corrected inner product coming from CFT is then recovered 
with the help of the supergravity interactions. At the factorization threshold, the leading large $N$ 
overlap is not vanishing; it is order one. So a Fock space structure with single gravitons 
corresponding to single traces, being orthogonal to multi-gravitons corresponding to
multi-traces, cannot be  the right spacetime structure for computing the leading large $N$ behaviour of the 
correlator. There should be a modification of 
the spacetime effective field theory which reproduces the correlators at threshold. This modification is unknown, but 
hints about its nature can be obtained by studying the detailed properties of the threshold. 

Once the angular momenta $J$ are sufficiently large that we are well past the threshold and into 
the region of broken factorization, we eventually reach the region of $J \approx N$, where 
the best way to think about the physics is in terms of giant gravitons \cite{bbns}. The basis of Schur polynomial 
operators, which are non-trivial linear combinations of multi-traces, becomes the best way to 
match bulk states and CFT states \cite{cjr}. The region where $ J \approx N$ was indeed earlier 
identified as an interesting region in connection with the fact that finite $N$ relations 
allow single traces to be expressed in terms of multi-traces via Cayley-Hamilton relations \cite{malstrom}.
This lead to a stringy exclusion principle, suggestive of some form of algebraic deformation 
of the spacetime algebra of functions \cite{jevram}. 

Here we focus instead on the threshold near $J \approx \sqrt { N \log N }$, where the large $N$ 
correlator is not infinite, but fixed at $G_3 = 1 $.  We could even take $G_3 = c$ for a small $c$, say 
$10^{-5}$,  but not going to zero as $N$ approaches infinity. So it is very plausible that a spacetime picture 
in terms of elementary objects whose number matches the number of traces, e.g. gravitons 
or gravitons stretched into BMN strings, is the right framework for understanding the precise nature of 
the threshold and the form of the interactions in this threshold region. 

With these motivations spelt out, we turn to some qualitative outcomes of our detailed studies 
of how the thresholds are approached when various parameters in the graviton system are tuned. 
An intriguing result we find is that, as we explain further in the next subsection, 
 in some of its  effects on the factorization threshold, separation in 
$J$-space is similar to separation in coordinate space. It is tempting to interpret this by associating the $J$-quantum 
number of a graviton to the radial AdS dimension, in the spirit of the UV-IR relation \cite{SusskWitt, peetpol}. 
This line of argument was adopted in the first version of the paper. This turns out to be rather subtle.\footnote{We thank 
the JHEP referees for comments on this point.} It is true that we can make an argument relating spatial extents to 
graviton energies by considering the LLM picture \cite{LLM}. The trace is a superposition of Schur polynomials
corresponding to hook representations, interpolating between a single row and a single column Young diagram. 
This is a superposition of states in the free fermion picture involving excitation of a fermion from some depth 
$k$ below the top of the Fermi sea to a level $(J - k)$ above the Fermi sea, with $k$ varying from $0$ to $(J-1)$.  
Since the fermion energy levels translate to radial positions in the LLM plane, with large radial positions of  the excited fermion being 
closer to the boundary, this is in line with the naive UV-IR argument.
However, consideration of normalizable modes in the global coordinates 
shows that gravitons at higher energy $J$ become more localized near the centre \cite{maldacenagg}. 
This suggests that the interpretation of half-BPS correlators in terms of gravitons 
requires care regarding the distinction between  normalizable and non-normalizable modes of the same field, and between 
the Lorentzian versus Euclidean picture of AdS. 
It is therefore prudent to postpone a detailed spacetime 
interpretation of the thresholds at this stage. It is nevertheless clear that this 
breakdown of the standard Fock space structure of effective spacetime field theory is an important new 
window where the gauge theory can provide valuable information 
to guide the spacetime understanding.

\subsection{Refined investigations of the factorization thresholds} 
In Section \ref{sec:j1neqj2} we investigate the more general extremal normalized three-point correlator
\bea
G_3(J_1, J_2, N) = \corb{\tr Z^{J_1}\tr Z^{J_2} \tr Z^{\dagger J_1+J_2}}
\eea
where $J_1 \neq J_2$. We define the threshold to be the surface in the three-dimensional parameter space $(J_1, J_2, N)$ that satisfies
\bea
G_3(J_1, J_2, N) = 1. \label{eq:thresholdsurface}
\eea
Making the assumption that both $J_1$ and $J_2$ grow at least as large as a positive power of $N$, then we find in
Section \ref{sec:j1neqj2} that
the correlator decays to zero if the product of the angular momenta $J_1 J_2$ is less than $N$ at large $N$, 
and grows exponentially if $J_1J_2$ grows faster than $N^{1+\delta}$ with $N$, where $\delta$ is any positive constant.
If the angular momenta are constrained to lie in the range $N< J_1J_2 < N^{1+\delta}$, then an asymptotic 
form of the correlator can be found. We find that at large $N$ in this regime, the threshold lies at
\bea
J_1  J_2 \approx N \log N,
\eea
where we have dropped a constant multiplicative factor.

\begin{figure}[t]
\centering
\includegraphics[width=0.6\textwidth]{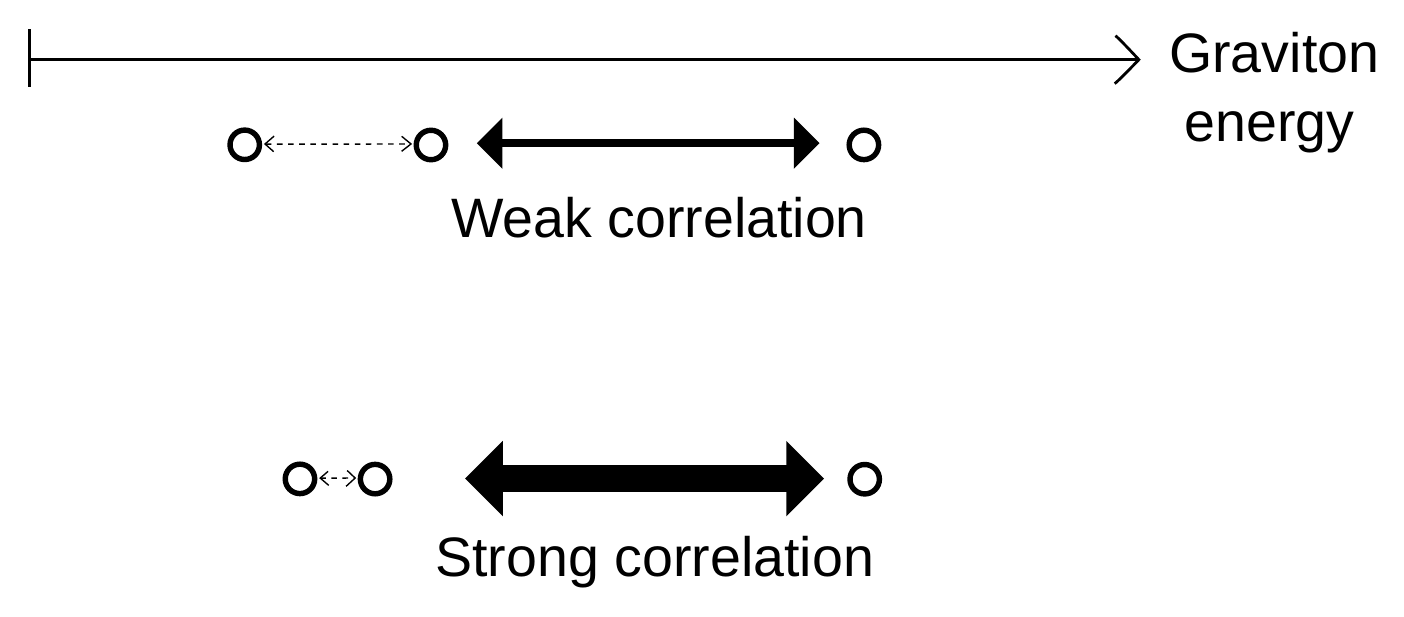}
\caption{
Two systems of gravitons with different energy differences but the same total energy.
Graviton states become strongly correlated when the separation of the graviton energies decreases.  }
\label{fig:bulkseparation}
\end{figure}

In the bulk picture, single trace operators with different dimensions correspond to gravitons at different energies. 
The combined energy of the two gravitons with energies $J_1$ and $J_2$ is equal to the energy of the other graviton $(J_1+J_2)$.
If we fix $N$ and the energy $(J_1+J_2)$ of the more energetic graviton, but vary the {\it difference} in the energies of the less
energetic gravitons $\Delta J = |J_1-J_2|$, 
then we find that we can move within parameter space \emph{from the factorization region to the threshold 
by decreasing the difference in energies of the two gravitons}. This is illustrated in Figure \ref{fig:bulkseparation}.

\begin{figure}[t]
\centering
\includegraphics[width=0.6\textwidth]{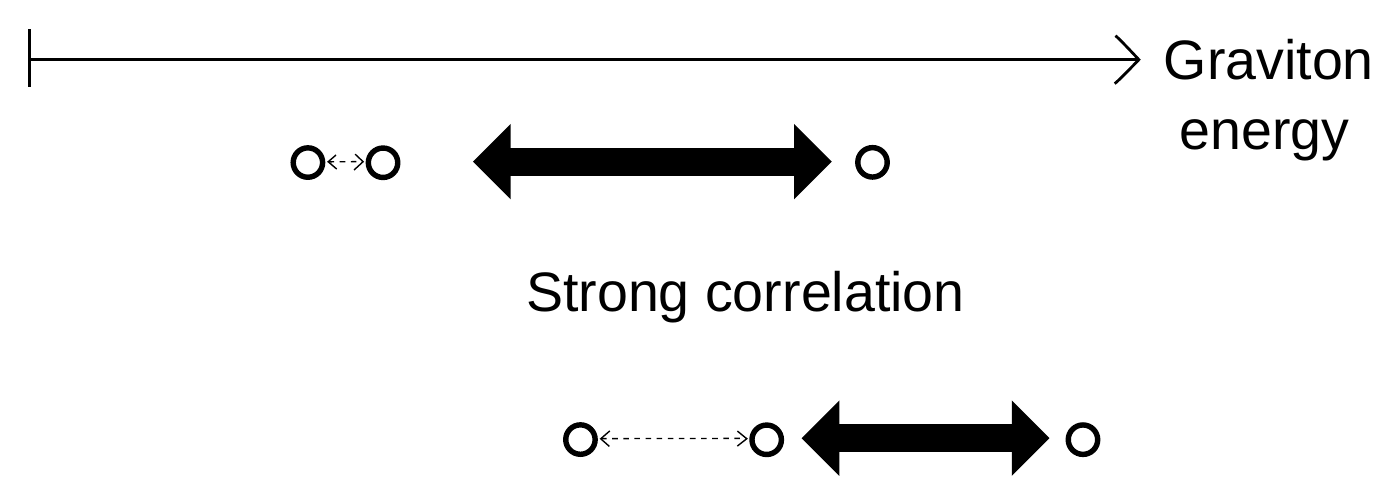}
\caption{
Two systems of gravitons at the threshold with different energy separations.
Graviton states become strongly correlated at lower energies (further from the boundary) when the separation of the graviton energies is smaller. }
\label{fig:bulkfactorization}
\end{figure}

Another related set-up is a strongly-correlated system of gravitons at the threshold in which
$N$ and the value of the correlator $G_3(J_1, J_2, N)=1$ are fixed but the separation of the graviton 
energies is varied.
Once $N$ is fixed and we are constrained to the threshold surface, there is only one available free 
parameter in the system, which we take to be the separation of the graviton energies $|J_1-J_2|$.
It can be shown that \emph{increasing the separation in energies $|J_1-J_2|$ of the two gravitons
at the threshold corresponds to an increase in the energy $(J_1+J_2)$ of the single graviton state.}
This system is shown in Figure \ref{fig:bulkfactorization}.   

We extend the investigation of factorization thresholds to the case of non-extremal correlators. 
In particular we study in detail the multiparticle-normalized correlator
\bea 
\corb{\str(Z^{J_1}Y)\str(Z^{J_2}Y^{\dagger})\tr(Z^{\dagger J_1+J_2})}
\eea 
and find a sensible extension of the discussion of factorization thresholds from the extremal case. 
In the discussion of extremal correlators above, we did not 
pay much attention to the spatial dependences of correlators. There is a simple reason for this. 
In the extremal case, we can set the two holomorphic operators at one point $x_1$ and the 
anti-holomorphic operator at another point $x_2$. This has the standard dependence 
$|x_1 - x_2|^{ - 2 (J_1 + J_2 ) }$. The spatial dependence can be removed by taking the anti-holomorphic 
operator to infinity, changing frame by the inversion $y  = \frac{x_2}{|x_2|^2}$. In this limit 
the correlator is computing an inner product of states and all position dependences disappear after we take into account 
the conformal transformation of the anti-holomorphic operator. In the above non-extremal case we can set the first operator at $x_1$, the second at $x_1 + \epsilon$ and take the third operator to infinity by applying an inversion. 
The only position dependence left is $\epsilon^{ -2}$. So the above correlator is a dimensionful quantity 
and it does not make sense to ask when it is equal to one in the large $N$ limit. 

We can introduce a dimensionful energy cutoff $\Lambda$ in the CFT. 
This dimensional cutoff will not change the CFT calculation if we take $\epsilon \Lambda \gg 1$.
The correct quantity to use to define the threshold is then $\Lambda^{-2}$ times the non-extremal correlator above.
This will be dimensionless, will contain the dimensionless parameter $\epsilon \Lambda \equiv \cR$ and can be compared to one
to define a factorization threshold. In the region of $J_i$ of order one and $ \cR \approx 1$  there is factorization,
but appropriate growth of $J_i$ with $N$ can cause breakdown of factorization, the details of the threshold 
depending on the dimensionless $ \cR$. 
We find that \emph{decreasing $ \epsilon \Lambda $, within the regimes where the correlator calcuations are valid, 
can cause the transition from factorization to breakdown}. This is in line with the discussion in 
\cite{PapRaju}, where short distances were argued to explore large energies which have to be low enough 
in relation to $N$ for factorization to hold.  
Another interesting aspect of this nearly-extremal correlator is that 
when $\cR=\epsilon \Lambda$ is large and fixed, or only varies with $N$ as a power or less,
then the threshold is of the same form as the extremal correlator; we find 
the threshold lies at $J_1J_2 \approx N\log N$.

Later in the paper, we consider the transition from multiple holomorphic traces 
to a single anti-holomorphic trace, or equivalently multiple gravitons going to a single 
graviton. If we have $k$ starting gravitons, with $k$ order $1$, we find that the threshold 
depends on the \emph{largest pairwise product} $J_i J_j$, and occurs at 
$ J_i J_j \sim k^{-1} N \log N$. The threshold of factorization 
\emph{decreases} as the number of gravitons in the multi-graviton state \emph{increases}.

Another generalization of the threshold investigation involves 
considering three-point extremal correlators corresponding to graviton scatterings on an LLM background given
by $M$ maximal giant gravitons, as in \cite{dmk09}. When $M$ is of the same order as $N$, then the factorization threshold
is $J_1 J_2 \approx (M+N)\log (M+N)$. If $M$ is chosen to have a fixed linear dependence on $N$, then the leading order behaviour
of the threshold is again $J_1 J_2 \approx N \log N$, up to a constant factor.

We conclude that another striking property of the thresholds is the 
universality of the leading large $N$ behaviour of the form $J_i J_j \sim N \log N$. 

\section{The extremal three-point correlator with \texorpdfstring{$J_1=J_2$}{J1=J2}}\label{sec:j1eqj2}
In this section we present a detailed calculation of the asymptotic form of the three-point correlator
\bea
G_3(J, N) = \corb{\tr Z^J \tr Z^J \tr Z^{\dagger 2J}} \label{eq:corb}
\eea
in the relevant region $N^\frac{1}{2}< J < N^{\frac{1}{2}+\delta}$, where $\delta$ is any small positive constant.
We then asymptotically solve the threshold equation
\bea
G_3(J_t(N), N)= 1 \label{eq:threshold}
\eea
in the large $N$ limit, deducing that at leading order the threshold behaves as
\bea
J_t \sim \sqrt{\frac{N}{2}\log N}.
\eea
Further, we find explicitly the all-orders asymptotic expansion of the threshold, and attempt to extend this result past perturbation theory
by deriving a transseries expansion. Finally, we discuss some links between the form of the threshold solution and running couplings in QCD.

\subsection{Review of asymptotics and series} \label{sec:series}
We start by briefly reviewing and clarifying some definitions, and introducing some new notation.
Throughout this paper, we will be using the precise mathematical definition of the asymptotic symbol `$\sim$',
the `little o' order symbol $o$, and asymptotic series.
We will also be using a precise definition of the `big O' order symbol $\cO$ that differs slightly from that used in the literature, 
but which is stronger than the commonly-used definition. 

For two $N$-dependent functions $f(N)$ and $g(N)$, then we say that $f\sim g$ at large $N$ if 
\bea
\lim_{N\to\infty} \frac{f(N)}{g(N)} = 1.
\eea
Note that with this definition the ratio of these two functions must tend to one, and not to any other constant.
We use the notation $f = o(g)$ if $f$ is a function that satisfies
\bea
\lim_{N\to\infty}\frac{f(N)}{g(N)} = 0,
\eea
i.e. if $f$ is much smaller than $g$ at large $N$. 
From these definitions, the following two statements are equivalent:
\bea
f(N) &\sim & g(N) \ret
f(N) &=& g(N)(1 + o(1)). \label{eq:asympequiv}
\eea
We shall also use the notation $f \ll g$ if $f = o(g)$, and conversely $f \gg g$ if $g = o(f)$.
An {\it asymptotic series} at large $N$ is formally defined by a set of functions $\{\phi_k(N)\}$ and constant coefficients $\{a_k\}$ with the property that 
\bea
\phi_{k+1} = o(\phi_{k})
\eea
for any $k\geq 0$. We say that
\bea
f(N) \sim \sum_{k=0}^\infty a_k \phi_k
\eea
if, for any $n\geq 0$, we have
\bea
f - \sum_{k=0}^n a_k\phi_k = o(\phi_n).
\eea
This definition of an asymptotic series does not allow for terms which are subleading to all the $\phi_k$.
Later, we shall also employ an extended version of an asymptotic series called a \emph{transseries}.
This type of series contains extra terms that tend to zero faster than all terms in a classical asymptotic series, 
but can still be assigned meaning when considered as a formal sum.
Transseries are commonly used in describing instanton corrections to series expansions
generated in QFTs, in which the instanton-dependent terms are exponentially suppressed in the coupling constant.
We discuss this more in Section \ref{sec:transseries}.

In this paper, we write $f=\cO(g)$ (or occasionally  $ f \approx g $) if there exists some positive constant $C$ such that
\bea
\lim_{N\to\infty}\frac{|f(N)|}{|g(N)|} = C.
\eea
This is a departure from the  $ \cO $ (big O)  notation in common use which only requires the ratio $f/g$ to be bounded from above at large $N$. 
This modified definition is a stronger condition as it not only implies that $f/g$ is bounded from above, 
but is also bounded from below too.
This is useful for keeping track of the errors and assumptions made at each step within our calculations.

The $\cO$  notation is used for expressing the errors of an $N$-dependent function, or corrections to an asymptotic series, or for giving
a coarse expression of the leading-order behaviour of a function.
It is used in the following for representing functions whose explicit forms are unknown or irrelevant, but whose leading-order behaviours
at large $N$ are important. 
Generally, when an upper bound on the leading-order behaviour of a correction is known but a lower bound is not, then we will use the  $ o $  (little o) symbol.
In general, we shall write equations as equalities when the corrections or errors are present, and use `$\sim$' for equations when the error terms have been dropped.

\subsection{Asymptotics of the three-point correlator}\label{sec:asymptotics}

To solve the threshold equation \refb{eq:threshold}, we need to find an asymptotic form of the normalized correlator \refb{eq:corb} at large $N$ and large $J$, with small $J/N$. 
The form of this expression will change depending on how quickly $J$ grows with $N$, so it is necessary to carefully specify at each stage what possible behaviour $J$ can take.
We will find that the breakdown threshold is located at $J$ just larger than $\cO(\sqrt{N})$, and so we will look for a large $N$ asymptotic form of the correlator $G_3(J, N)$ that is valid in this region. 
It suffices to impose $J\ll N^{2/3}$ to describe the asymptotic form of the correlator around the threshold.

The position-independent two-particle and three-particle correlators are known precisely for finite $N$ \cite{kpss}.
We recall that the two-point function at zero coupling is 
\bea
\cor{\tr Z^J \tr Z^{\dagger J}} = J!\left[ \binom{N+J}{J+1} - \binom{N}{J+1}\right], \label{eq:2pt}
\eea
and the three-point function (for general operator dimensions $J_1$ and $J_2$) is
\ml{
\cor{\tr Z^{J_1}\tr Z^{J_2} \tr Z^{\dagger J_1+J_2}} = (J_1+J_2)!\left[ \binom{N+J_1+J_2}{J_1+J_2+1} - \binom{N+J_1}{J_1+J_2+1} \right. \\ 
\left. - \binom{N+J_2}{J_1+J_2+1} + \binom{N}{J_1+J_2+1}\right], \label{eq:3pt}
}
All the terms in the finite $N$ correlator expressions are of the form
\bea
J!\binom{N+\Lambda}{J+1}  = \frac{(N+\Lambda-J)}{(J+1)} \frac{(N+\Lambda)!}{(N+\Lambda-J)!},
\eea
where $\Lambda$ is either $0$ or $J$ for the terms in the two-point function.
Taking $N$ and $J$ to be large, but keeping $J/N$ small, we apply Stirling's approximation
\bea
n! = e^{-n}n^{n+\frac{1}{2}}\sqrt{2\pi}\left(1+\hot{1}{n} \right)
\eea
to find that
\bea
J!\binom{N+\Lambda}{J+1} \sim \frac{(N+\Lambda - J)}{J+1}N^{J}e^{-J}\sqrt{1+\frac{\Lambda}{N}}\sqrt{1+\frac{\Lambda-J}{N}}
\left(1+ \frac{\Lambda}{N}\right)^{N+\Lambda}\left(1+ \frac{\Lambda-J}{N}\right)^{-N-\Lambda+J} \eea
\bea
\sim \frac{N^{J+1}e^{-J}}{J} \left(1+ \frac{\Lambda}{N}\right)^{N+\Lambda}\left(1+ \frac{\Lambda-J}{N}\right)^{-N-\Lambda+J}. \label{eq:bracketedfactor}
\eea
Here, we have dropped some error terms of order $\hot{1}{J}$ and $\hot{J}{N}$. We expand the terms in the brackets by taking logs, and using the fact that $\Lambda<N$ to perform a series expansion. We find that 
\bea
\log \left(1+\frac{\Lambda}{N}\right)^{N + \Lambda} &=&  -N\left(1+\frac{\Lambda}{N}\right)\sum_{k=1}^\infty \frac{(-\Lambda)^k}{k N^k} \\
&=& \Lambda + \sum_{k=1}^\infty \frac{(-\Lambda)^{k+1}}{k(k+1)N^k}.
\eea
Hence, replacing $\Lambda$ with $\Lambda-J$ in the second bracketed factor of \refb{eq:bracketedfactor}, we find
\bea
J!\binom{N+\Lambda}{J+1} \sim \frac{N^{J+1}}{J} \exp\left( \sum_{k=1}^\infty \frac{(-\Lambda)^{k+1} - (-\Lambda+J)^{k+1}}{k(k+1)N^k}\right). \label{eq:binomial}
\eea

We can simplify this expression by dropping the terms in the infinite sum that tend to zero with large $N$. The $k$th term in the sum scales like $J^{k+1}/N^k$ for some integer $k$, so if we impose that $J\ll N^{2/3}$, then all terms with $k\geq 2$ are small. With this condition, we can drop the subleading terms of order $\cO(J^3/N^2)$ and write 
\boxeq{
J!\binom{N+\Lambda}{J+1} \sim \frac{N^{J+1}}{J} \exp\left( \frac{-J^2}{2N} + \frac{J\Lambda}{N} \right). \label{eq:infsumdrop}
}
This expression, which is valid for any $\Lambda\leq J \ll N^{2/3}$, is used repeatedly in the following sections to derive the 
asymptotics of finite $N$ correlators. 
Including both terms in \refb{eq:2pt} with $\Lambda=J$ and $\Lambda=0$ respectively, we can now state 
that two-point function has the asymptotic form
\bea
\cor{\tr Z^J \tr Z^{\dagger J}} \sim \frac{N^{J+1}}{J}e^{\frac{J^2}{2N}}\left(1 - e^{-\frac{J^2}{N}}\right).
\eea

This approach generalizes in a straightforward manner to the three-point function. Replacing $J$ with $2J$ and allowing $\Lambda$ to take the values $0$, $J$, and $2J$, we find that \refb{eq:3pt} becomes
\bea
\cor{\tr Z^{J}\tr Z^{J} \tr Z^{\dagger 2J}} &\sim & \frac{N^{2J+1}}{2J}\left( e^{\frac{2J^2}{N}} - 2 + e^{\frac{-2J^2}{N}} \right) \\
&\sim & \frac{N^{2J+1}}{2J} e^{\frac{2J^2}{N}} \left( 1 - e^{-\frac{2J^2}{N}}\right)^2.
\eea
These expressions allow us to read off the asymptotic form of the normalized three-point function \refb{eq:corb}. We find that
\bea
G_3 \sim \sqrt{\frac{J}{2N}}  \exp\left(\frac{J^2}{2N}\right) \frac{\left(1-e^{-\frac{2J^2}{N}}\right)^2}{(1-e^{-\frac{J^2}{N}})\sqrt{(1-e^{-\frac{4J^2}{N}})} }. \label{eq:convoluted}
\eea
This expression is valid for any behaviour of $J$ provided that $J\ll N^{2/3}$. 

To find a more tractable version of this formula at large $N$, we need to state how $J^2/N$ grows with $N$.
There are three cases to consider: $J^2/N$ going to zero with large $N$, $J^2/N$ going to a constant, and $J^2/N$ going to infinity.
In the first case where $J^2/N$ is small, we can use
\bea
(1-e^{-\frac{kJ^2}{N}}) \sim \frac{kJ^2}{N}, \qquad \exp\left(\frac{J^2}{2N}\right) \sim 1
\eea
where $k \in \{1,2,4\}$, to see that
\bea
G_3 \sim \frac{\sqrt{J J (2J)}}{N},
\eea
which is the known behaviour of the normalized three-point correlator for $J\ll \sqrt{N}$. The assumption $J^2/N\to 0$ means that the correlator will tend to zero in this limit, and so factorization holds in this case.
Alternatively, in the case that $J^2/N$ tends to a constant value, i.e. $J=\cO(\sqrt{N})$, then \refb{eq:convoluted} will scale as $\cO(N^{-\frac{1}{4}})$ with large $N$. This means that factorization will still hold in this case.
However, in the case that $J^2/N$ grows large with $N$, then we have
\bea
(1-e^{-\frac{kJ^2}{N}}) \sim 1, \qquad  \exp\left(\frac{J^2}{2N}\right) \to \infty,
\eea
and thus
\boxeq{
G_3 = \corb{\tr Z^{J}\tr Z^{J} \tr Z^{\dagger 2J}} \sim \sqrt{\frac{J}{2N}}  \exp\left(\frac{J^2}{2N}\right). \label{eq:corthreshold}
}
This correlator will grow to infinity if $J$ grows quickly enough with $N$. In particular, if $J \geq N^{\frac{1}{2} + \delta}$ for some small constant $\delta>0$ at large enough $N$ i.e. if $J$ grows faster than $\sqrt{N}$ by a positive power, then the exponential term dominates and the correlator will tend to infinity. 
We deduce that the threshold - that is, the growth of $J$ with $N$ 
which keeps the correlator finite and non-zero at large $N$ - lies in the range 
\bea
N^\frac{1}{2} < J < N^{\frac{1}{2}+\delta}, \label{eq:region}
\eea
where $\delta$ is any small positive number.
This is the relevant region for solving asymptotically the factorization threshold equation
\bea
G_3(J_t(N), N) = 1.\label{eq:threshold2}
\eea

\subsection{Solving the factorization threshold equation}\label{sec:j1eqj2soln}

We can use \refb{eq:corthreshold} in the region \refb{eq:region} to find a function $J(N)$ that solves the threshold equation \refb{eq:threshold2} at large $N$. To do this, we write down the \emph{exact} equation
\bea
G_3 = \sqrt{\frac{J}{2N}}\exp\left(\frac{J^2}{2N}\right)e^{-\frac{1}{4}r}, \label{eq:r}
\eea
where the error function $r(J, N)$ is implicitly defined by this equation (the factor of $\frac{1}{4}$ here is chosen for later convenience).
All the large $N$ approximations that were taken in generating the asymptotic expression \refb{eq:corthreshold} 
are encoded in this error function, so it must tend to zero with $N$ (provided that we remain in the range \refb{eq:region}).
To find the leading-order behaviour of $r$, we collate the terms dropped at various stages in the previous section.
In \refb{eq:bracketedfactor} and  \refb{eq:infsumdrop}, we have dropped terms of order $\hot{1}{J}$, $\hot{J}{N}$, and $\hot{J^3}{N^2}$.
As $J^2/N$ is large, all these errors are $\hot{J^3}{N^2}$. Also, in performing the approximation 
\bea \left(1- e^{-\frac{kJ^2}{N}}\right) \sim 1 \eea
for various values of $k$, we have dropped terms of order $\cO(e^{-\frac{J^2}{N}})$. At present, we have not specified tight enough constraints on $J$ to determine which is the larger, so we keep both remainders. We write
\bea
G_3 = \sqrt{\frac{J}{2N}}\exp\left(\frac{J^2}{2N}\right)\left(1+ \hot{J^3}{N^2} + \cO\left(e^{-\frac{J^2}{N}}\right)\right)
\eea
and so we have
\bea
e^{-\frac{1}{4}r} = 1+ \hot{J^3}{N^2} + \cO\left(e^{-\frac{J^2}{N}}\right)
\eea
This means that the error function $r$ is bounded by
\bea
r = \hot{J^3}{N^2} +\cO\left(e^{-\frac{J^2}{N}}\right). \label{eq:errorfnbounds}
\eea
Again, we know that this function tends to zero, 
but can't yet deduce its leading-order behaviour before solving the threshold equation.
Rearranging \refb{eq:r}, we can write the threshold equation $G_3(J_t(N), N)=1$ as
\bea
\left(\frac{2J_t^2}{N}\exp\left({\frac{2J_t^2}{N}}\right)\frac{1}{8Ne^r} \right)^\frac{1}{4} = 1. \label{eq:rearranged}
\eea
This equation cannot be solved exactly in terms of elementary functions (e.g. exponentials, logarithms and powers of $z$), 
but it can be rewritten and approximated by using the Lambert $W$-function. 
The Lambert $W$-function is defined by the equation 
\bea
W(z)e^{W(z)}= z.
\eea
It is a multivalued function, but here we just consider the principle branch of the function, where $W(z)$ is positive and real for positive real $z$.
In this regime, a large $z$ asymptotic expansion of the function is known to all orders \cite{w96, w97}.
Equation \refb{eq:rearranged} is solved in terms of the $W$-function by
\bea
\frac{2J_t^2}{N} = W(8Ne^{r}),
\eea
which can be written
\bea
J_t =  \sqrt{\frac{N}{2}W(8Ne^{r})}. \label{eq:wsolution}
\eea

To find a more tractable version of the threshold expressed in terms of elementary functions, we can expand the $W$-function by using
its asymptotic series.
The large $z$ expansion of the $W$-function is \cite{w97}
\bea
W(z) \sim \log z - \log \log z +\sum_{n=1}^\infty \left(\frac{-1}{\log z}\right)^n \sum_{k=0}^n \stirling{n}{n - k + 1} \frac{(-\log \log z)^k}{k!}, \label{eq:wseries}
\eea
where the coefficients in the square brackets are the Stirling cycle numbers (of the first kind); the notation $\stirling{n}{k}$ denotes the number of permutations of $n$ elements composed of $k$ disjoint cycles. 
We can find the leading-order behaviour of the threshold by truncating this series. However, to guarantee that 
the truncated solution still satisfies $G_3(J_t(N),N)=1$ in the large $N$ limit, we need to keep all the terms in the series 
that do not tend to zero.
The first two terms in the series are large as $z\to\infty$, and the remaining terms in the infinite series all go to zero, and so we keep the first two terms and find that the large $N$ solution of \refb{eq:wsolution} is
\bea
\frac{J_t^2}{N} = \frac{1}{2}\left[\log (8Ne^{r}) - \log \log (8Ne^{r}) + \hot{\log \log N}{\log N}\right].
\eea
We can now extract out the $N$-dependence of the remainder function at the threshold, $r(J_t(N), N)$. Since
\bea
\frac{J_t^2}{N} = \frac{1}{2}\left(\log 8N - \log \log N + o(1) \right),
\eea
we find that 
\bea
e^{-\frac{J_t^2}{N}} \sim \sqrt{\frac{\log N}{8N}}, \qquad \frac{J_t^3}{N^2} \sim \sqrt{\frac{(\log N)^3}{8N}},
\eea
and so to leading order in $N$,
\bea
r(J_t(N), N) = \cO\left(\sqrt{\frac{(\log N)^3}{N}}\right).
\eea
This term is smaller than $N^{-\frac{1}{2}+\delta}$ for any constant $0<\delta<\frac{1}{2}$, and so all powers of $r$ are 
subleading to all logarithm-dependent terms in the expansion. We can therefore discard these $r$-dependent terms as they are
`exponentially suppressed' in terms of the parameter $\log N$.
The full asymptotic series expansion of the threshold is
\bea
J_t^2 \sim \frac{1}{2}N \left[ \log (8N) - \log \log (8N) +\sum_{n=1}^\infty \left(\frac{-1}{\log (8N)}\right)^n \sum_{k=0}^n \stirling{n}{n - k + 1} \frac{(-\log \log (8N))^k}{k!} \right]. \quad  \label{eq:fullseries}
\eea
Taking square roots and moving out the constant factors in the logs, we deduce that the leading-order terms in the expansion of the threshold are
\bea
J_t = \sqrt{\frac{1}{2}N\log N}\left[1 - \frac{\log \log N}{2\log N} + \frac{\log 8}{2\log N} + \hot{(\log \log N)^2}{(\log N)^2}\right]. \label{eq:thresholda}
\eea
This is the leading-order solution to 
\bea
G_3(J_t(N), N) := \corb{\tr Z^{J_t} \tr Z^{J_t} \tr Z^{2J_t}}=1
\eea
for large $N$ and large $J_t$. 

In \refb{eq:thresholda}, we have given the first three terms in the expansion of the threshold. This is the necessary degree of accuracy
of the threshold $J_t(N)$ for which the truncated series still satisfies the threshold equation in the large $N$ limit. That is, if we take
the truncated threshold
\bea
\tilde{J}(N) = \sqrt{\frac{1}{2}N\log N}\left[1 - \frac{\log \log N}{2\log N} + \frac{\log 8}{2\log N}\right]
\eea
and plug this into the exact expression \refb{eq:r}, we have
\bea
G_3(\tilde{J}(N), N) = \exp\left[\frac{1}{16\log N}\left(\log \left(\frac{8}{\log N}\right)\right)^2 - \frac{1}{4}r\right],
\eea
which tends to one in the large $N$ limit. If we had only taken the first term in the threshold solution $\tilde{J}= \sqrt{\frac{1}{2}N \log N}$ 
and plugged this into \refb{eq:r}, we would have found that $G_3(\tilde{J}(N), N)$ actually grows logarithmically with $N$, and so
the threshold equation cannot hold for arbitrarily large $N$.
Similarly, truncating the series at the second term causes the correlator $G_3(\tilde{J}(N), N)$ to converge to a different constant than 1
at large $N$.

We remark that the factors of 8 appearing in the logs have come from choosing the factorization threshold to be at $G_3=1$. If we had instead chosen $G_3(J, N)= c$ for some constant $c$, then the threshold solution would be
\bea
\frac{J_t^2}{N} = \frac{1}{2} W(8c^4Ne^{r}), \label{eq:exactlambda}
\eea
and the leading-order behaviour after expansion would be
\bea
J_t = \sqrt{\frac{1}{2}N\log N}\left[1 - \frac{\log \log N}{2\log N} + \frac{\log 8 + 4\log c}{2\log N} + \hot{\log \log N}{\log N}^2 \right]. \label{eq:thresholdb}
\eea

\subsection{Similarities to the running coupling of non-abelian gauge theories } 
We pause here to discuss some similarities between our threshold solution and the the running coupling of non-abelian gauge theories. 
The beta function of $\alpha_s(Q^2)$ from QCD gauge theory is
\bea 
Q^2 \frac{ d \alpha_s  }{ d Q^2  } = \frac{d\alpha_s}{d L} = \beta_0 \alpha_s^2 + \beta_1 \alpha_s^3 + \beta_2 \alpha_s^3 + \cO(\alpha_s^4),
\eea
where $Q^2$ is the energy scale, $\beta_i$ is the beta function at loop order $(i+1)$, and $L  = \log ( \frac{ Q^2 }{ \Lambda^2 } )$. 
This has been solved perturbatively \cite{Czakon:2004bu, Alekseev} for the running coupling $\alpha_s ( Q^2 )$, 
\begin{multline}
\alpha_s ( Q^2 )  + \frac{1  }{ \beta_0  \log L }  = \frac{ \beta_1 }{\beta_0^2 L  } ( \log L ) + \frac{ \beta_1^2 }{ \beta_0^4 L^2  } \left ( 
( \log L)^2 - \log L  -1 + \frac{  \beta_0 \beta_2 }{ \beta_1^2 }  \right ) \\
+ \frac{ \beta_1^3 }{ \beta_0^6 L^3 } \left (  ( \log L )^3  - \frac{ 5 }{ 2 }  (\log L )^2  -
 ( 2  -  3 \frac{ \beta_0 \beta_2 }{ \beta_1^2 } ) \log L  +\frac{ 1  }{ 2 }  - \frac{ \beta_0^2 \beta_3  }{ \beta_1^3 } \right ) + \hot{(\log L)^4}{L^4}. \label{eq:QCDcoupling}
\end{multline}

The threshold solution can be recast into a form which reveals a striking similarity with the expansion of
$\alpha_s(Q^2)$.
Starting from the definition of the $W$-function and its asymptotic series \refb{eq:wseries}, we can write
\bea
\log W(z) &=& \log z - W(z) \\ &\sim&  \log \log z - \sum_{n=1}^\infty \left(\frac{-1}{\log z}\right)^n \sum_{k=0}^n \stirling{n}{n - k + 1}\frac{(-\log \log z)^k}{k!}, \label{eq:logofw}\eea
where the factors $\stirling{n}{k}$ are Stirling cycle numbers of the first kind.
Introducing the new variables $y= \log J_t$ and $v= \log N$, we can take logs of the exact solution 
\bea
J_t =  \sqrt{\frac{N}{2}W(8Ne^{r})}
\eea
and plug in the first few Stirling numbers to find 
\ml{
2y = v + \log v - \log 2 + \frac{1}{v}\left( - \log v + \log 8 \right) \\
+ \frac{1}{v^2}\left[ -\frac{1}{2}(\log v)^2 + (1+ \log 8)\log v - \frac{1}{2}(\log 8)(\log 8 + 2) \right] +  \cO\left( \frac{(\log v)^3}{v^3}\right) \label{eq:vseries}
}
\bea 
\sim v + \log v - \log 2 + \sum_{l=1}^\infty \frac{P^l_0(\log v)}{v^l},
\eea
where $P^l_0$ are polynomials of order $l$, and we have dropped the subleading $r$-dependent terms. All but the first three terms in this sum tend to zero in the large $v$ (i.e. large $N$) limit, so we can define the variable $Y = 2 y - v - \log v + \log 2$, which has the perturbative expansion
\begin{multline}
Y = \frac{1}{v}\left( - \log v + \log 8 \right) + \frac{1}{v^2}\left[ -\frac{1}{2}(\log v)^2 + (1+ \log 8)\log v - \frac{1}{2}(\log 8)(\log 8 + 2) \right] \\ +  \cO\left( \frac{(\log v)^3}{v^3}\right). \label{eq:logsimilarity}
\end{multline}

We can now see that both \refb{eq:QCDcoupling} and \refb{eq:logsimilarity} are manifestly of the same form.
Each bracketed term in the first series can be written $L^{-n}\tilde{P}^n(\log L)$, 
and each bracketed term in the second series can be written $v^{-n}P^n(\log v)$, where $\tilde{P}$ and $P$ are polynomials
of order $n$.
The similarity between these series is intriguing, and it would be of interest to find out if there is a physical explanation.

\subsection{Expansion of the threshold as a transseries} \label{sec:transseries}

We have given in equation \refb{eq:fullseries} an infinite asymptotic series expansion
of the threshold $J_t(N)$ in terms of powers of $\log N$ and $\log \log N$.
We can go beyond this classical asymptotic series approach to the threshold by considering the 
\emph{non-perturbative corrections}, generated by the subleading terms 
in $r$ that were previously neglected. This type of series is known as a \emph{transseries},
and is perhaps most commonly seen in theoretical physics to describe instanton corrections
in quantum field theory. 

When considering asymptotic expansions from path integrals in quantum field theory,
we are interested in not only the original perturbative series in the coupling constant, 
but also the exponentially-suppressed instanton correction terms.
These typically come from saddle-points in the path integral.
A typical asymptotic series in a quantum field theory with small coupling constant $g\to 0$
and instanton corrections has the form
\bea
\sum_n a_n g^{n} + e^{-A/g}\sum_{n}a_n^{(1)}g^{n} + \cO\left(e^{-2A/g}\right). \label{eq:instanton}
\eea
The definition of an asymptotic series given in Section \ref{sec:series} cannot 
be used to describe the exponential contributions, as they are subleading to all powers of the
coupling $g$. We make sense of a series with instanton corrections by thinking of it as 
a purely formal sum, in which $g$ and $e^{-A/g}$ are treated as independent variables. Once the formal 
transseries is constructed, there are approaches that can recover the \emph{exact} 
full form of the path integral from the series; this is called the
theory of \emph{resurgence}.
The lecture notes \cite{marino} give a review of transseries and resurgence in QFT and string theory. 

In our analysis of the threshold, the series we have found has not come from a path integral, but 
still has exponentially-suppressed corrections. Rather than corresponding to saddle-points, the exponential corrections
arise from the corrections to the asymptotics of the finite $N$ correlators.
We can see the analogy between thresholds and instanton expansions by changing variables from
$N$ to $v = \log N$ in our threshold expressions; the remainder term $r$ is then proportional to $e^{-v/2}$. We show in the following that
the general form of a transseries of the threshold can be found, in terms of $e^{-v/2}$, $v$ and $\log v$.

An interesting possible future research direction would be to use the transseries expansion to search for an effective field 
theory description of gravitons at the threshold. 
The threshold expansions with exponential corrections strongly resemble instanton expansions of field theoretic partition functions, 
and so they could well contain valuable hints about the nature of such an effective field theory.

We start by writing the threshold in terms of the variables $y=\log J_t$ and $v=\log N$ introduced in the previous section, but retain the $r$-dependent terms in the series expansion.
With the $r$-corrections, the series \refb{eq:vseries} becomes
\ml{ \label{eq:rcorrection}
2y = v + \log v - \log 2 + \frac{1}{v}\left( - \log v + \log 8 \right) + \frac{r}{v} + \hot{r^2}{v^2} \\
+ \frac{1}{v^2}\left[ -\frac{1}{2}(\log v)^2 + (1+ \log 8)\log v - \frac{1}{2}(\log 8)(\log 8 + 2) \right] +  \cO\left( \frac{(\log v)^3}{v^3}\right)
}
\bea 
\sim v + \log v - \log 2 + \sum_{l=1}^\infty \frac{P^l_0(\log v)}{v^l} + \hot{r}{v},
\eea
All the terms that depend on the error function $r$ are subleading to any power of $\log v$ and $v$. 
To find the exponentially-supressed contributions to the threshold and extend the asymptotic series to a transseries,
we need to find a more precise expression for $r$ near the threshold.
In the previous section, the function $r(J,N)$ was defined by the exact equation 
\bea
G_3 = \sqrt{\frac{J}{2N}}\exp\left(\frac{J^2}{2N}\right)e^{-\frac{1}{4}r}.
\eea
The next-to-leading order corrections to the remainder function $r$ were estimated in \refb{eq:errorfnbounds}.
A more careful calculation shows that the next-to-leading order behaviour of the correlator near the threshold is
\bea
G_3 = \sqrt{\frac{J}{2N}}\exp\left(\frac{J^2}{2N}\right)\left[ 1 - \frac{J^3}{N^2} + e^{-\frac{J^2}{N}}
+ \hot{J^6}{N^4} \right],
\eea
and so 
\bea
r(J,N) &=& 4\left( \frac{J^3}{N^2} - e^{-\frac{J^2}{N}} + \hot{J^6}{N^4}\right).
\eea
Plugging in the leading-order behaviour of the threshold $J\sim \sqrt{\frac{1}{2}N\log N}$ gives us the leading-order behaviour of $r$ as a function purely of $N$, or as a function of $v$. We find
\bea
r(J_t(N), N) &=& \sqrt{\frac{2(\log N)^3}{N}}\left[1 - \frac{1}{\log N}\left(\log \log N - \frac{3}{2}\log 8 + 1\right) + \hot{\log \log N}{\log N}^2 \right] \qquad \\
&=& \sqrt{2}v^{\frac{3}{2}}e^{-\frac{v}{2}}\left[1 - \frac{1}{v}\left(\log v - \frac{3}{2}\log 8 + 1\right) + \hot{\log v}{v}^2\right]
\eea
This correction can be reintroduced into \refb{eq:rcorrection} to give the first exponential
correction of the threshold,
\ml{
2y = v + \log v - \log 2 + \frac{1}{v}\left( - \log v + \log 8 \right) + \hot{\log v}{v}^2
\\
+ \sqrt{2ve^{-v}} \left[1 + \frac{1}{v}\left(\frac{3}{2}\log 8 - 2 - \log v\right) + \hot{\log v}{v}^2 \right] + \cO\left( v e^{-v}\right).
}

The remainder $r$ has an asymptotic expansion at the threshold as a series of powers of $v^\frac{3}{2}e^{-\frac{v}{2}}$ multiplied
by powers of $\log v$ and inverse powers of $v$.
From considering the structure of the terms in \refb{eq:logofw}, and writing $8Ne^r = e^{v+\log 8 + r}$, it can be seen that a $k$th power
of $r$ in the asymptotic expansion of $W(8Ne^r)$ is accompanied by a $k$th inverse power of $v$, followed by positive powers of
$\log v$ and inverse powers of $v$. 
Noting that the subleading terms in the asymptotic expansion of $r$ can also contribute, we can deduce the 
all-orders form of the asymptotic series with exponential corrections, although it is difficult to calculate 
coefficients explicitly beyond the first few terms. The general form of the transseries form of the threshold is
\bea
2y \sim v + \log v - \log 2 + \sum_{k=0}^\infty \sum_{n=0}^\infty  (\sqrt{v e^{-v}})^k \frac{P^n_k(\log v)}{v^{n}},
\eea
where the $P^n_k$ are polynomials of order $n$, and $P^0_0(\log v)=0$.

This series gives an alternative expression for the threshold $J_t = e^y$ in terms of $v=\log N$. 
Only the first three terms do not go to zero in the large $N$ limit, so we can exponentiate this expression to derive an infinite 
asymptotic series for the threshold. We find that
\bea
J_t  \sim \sqrt{\frac{1}{2}N\log N}\left[ 1 + \sum_{k=0}^\infty \sum_{n=0}^\infty \left(\sqrt{\frac{\log N}{N}}\right)^k \frac{P'^n_k(\log \log N)}{(\log N)^{n}}\right]
\eea
where the polynomials have been modified, but the form of the series has not. 
As remarked at the end of subsection \ref{sec:j1eqj2soln}, for a truncated threshold series $\tilde{J}_t(N)$ to satisfy $G_3(\tilde{J}_t(N),N)\to 1$ at large $N$, we must include the next-to-leading order term,
\bea
\frac{P'^1_0(\log \log N)}{\log N} = \frac{-\log \log N}{2\log N} + \frac{\log 8}{2\log N}.
\eea

As a final remark, we note again that changing the threshold from $G_3=1$ to $G_3=c$ will not alter the form of the series, but will
modify the polynomials and constants. From \refb{eq:exactlambda}, we see that shifting the threshold equation to $G_3=c$ will transform the series as
\bea
&2y& \sim v + \log v - \log 2 + \sum_{k=0}^\infty \sum_{n=0}^\infty  (\sqrt{v+c^4}e^{-\frac{v+ c^4}{2}})^k \frac{P^n_k(\log (v + c^4))}{(v+c^4)^{n}}\ \qquad \\
&\sim &  v + \log v - \log 2 +  \sum_{k=0}^\infty \sum_{n=0}^\infty  (\sqrt{v}e^{-\frac{v}{2}})^k \frac{\tilde{P}^n_k(\log v)}{v^{n}}.
\eea
The three leading-order terms and the highest-order terms in the polynomials are unaffected by the shift.

\section{The extremal three-point correlator with \texorpdfstring{$J_1\neq J_2$}{J1=/=J2} }\label{sec:j1neqj2}
In the previous section we solved the equation 
$G_3(J(N), N) = 1$
at large $N$ by finding the asymptotic form of the three-point function $G_3$ and solving for $J(N)$. 
In this section we consider the threshold of factorization for the more general three-point function,
\bea
G_3(J_1, J_2, N) &:=& \corb{\tr Z^{J_1}\tr Z^{J_2} \tr Z^{\dagger J_1+J_2}} \ret
&=&\frac{ \cor{\tr Z^{J_1}\tr Z^{J_2}\tr Z^{\dagger J_1+J_2}} }{\sqrt{\cor{\tr Z^{J_1}\tr Z^{\dagger J_1}} \cor{\tr Z^{J_2}\tr Z^{\dagger J_2}} \cor{\tr Z^{J_1+J_2}\tr Z^{\dagger J_1+J_2}} }},
\eea 
and examine the behaviour of $J_1(N)$, $J_2(N)$ with $N$ for which the threshold equation 
\bea
G_3(J_1(N), J_2(N), N) = 1 \label{eq:thresholdj1j2}
\eea
is satisfied at large $N$.
Using similar methods as in the previous section, the asymptotic form of $G_3(J_1(N), J_2(N), N)$ can be found at large $N$,
and can be used to invert the threshold equation \refb{eq:thresholdj1j2} to retrieve a simple leading-order constraint
on the functions $J_1(N), J_2(N)$ at the threshold. We find quite generally that \refb{eq:thresholdj1j2}
is solved in the large $N$ limit by solutions $J_1(N), J_2(N)$ that have the leading-order behaviour
\bea
J_1J_2 \approx N \log N,
\eea
where we have omitted a constant of proportionality.
In fact, this constant of proportionality depends on the $N$-dependent behaviour of the \emph{smaller} 
of the two angular momenta $J_1$ and $J_2$.

In the following subsection, we present the calculation of the large $N$ behaviour of the correlator $G_3(J_1, J_2, N)$, and invert the
threshold equation $G_3(J_1(N), J_2(N), N)=1$ to find the result $J_1J_2 = \cO(N \log N)$. 
Following that, we discuss how the threshold from the bulk perspective relates the separation of the graviton
energies $\Delta J = |J_1 - J_2|$ to the energy of the single graviton $E=(J_1+J_2)$.

\subsection{Scaling limits and the threshold equation}

We start from the expressions for the two and three-point correlators in Section \ref{sec:asymptotics}. These generalize in a straightforward manner to give the expression, valid for large $N$ and $1 \ll J_1, J_2 \ll N^{\frac{2}{3}}$:
\bea
G_3(J_1, J_2, N) \sim \sqrt{\frac{J_1J_2}{(J_1+J_2)N}}\exp\left(\frac{J_1J_2}{2N}\right)\frac{(1-e^{-\frac{J_1(J_1+J_2)}{N}})(1-e^{-\frac{J_2(J_1+J_2)}{N}})}{\sqrt{(1-e^{-\frac{J_1^2}{N}})(1-e^{-\frac{J_2^2}{N}})(1-e^{-\frac{(J_1+J_2)^2}{N}})}}. \label{eq:3ptj1j2}
\eea
Without loss of generality, we assume throughout that $J_1\leq J_2$. 

We can find bounds on the threshold region by considering the large $N$ behaviour
of the product of the angular momenta $J_1J_2$.
If $J_1J_2/N$ goes to zero with $N$, then the assumption $J_1 \leq J_2$
means that $J_1^2/N$ must also go to zero with $N$.
We note that 
\bea
1-e^{-\frac{J_1^2}{N}} &\sim& \frac{J_1^2}{N}, \ret
1-e^{-\frac{J_1(J_1+J_2)}{N}} &\sim& \frac{J_1(J_1+J_2)}{N}, \ret
1-e^{-\frac{J_2(J_1+J_2)}{N}} &\sim& 1-e^{-\frac{J_2^2}{N}}, \ret
1-e^{-\frac{(J_1+J_2)^2}{N}} &\sim& 1-e^{-\frac{J_2^2}{N}}
\eea
to deduce that the the correlator behaves as
\bea
G_3 \sim \sqrt{\frac{J_1J_2}{N}}\sqrt{\frac{J_1+J_2}{N}} \ll 1.
\eea
The correlator thus decays to zero at large $N$.
On the other hand, if $J_1J_2/N$ grows with $N$ to infinity at a faster rate than some small positive power of $N$, 
i.e. $J_1J_2 \geq N^{1+\delta}$ for some small positive constant $\delta$,
then the $\exp(J_1J_2/2N)$ factor scales at least as quickly as $\exp(N^\delta)$, an exponential of a positive power of $N$.
All other factors in the expression are bounded by powers of $N$, and so the exponential term dominates
and $G_3$ must tend to infinity. 
Summarizing the above, we have
\boxeq{
G_3(J_1,J_2, N) \to 0,& \qquad \frac{J_1J_2}{N}\to 0, \ret
G_3(J_1,J_2, N) \to \infty,& \qquad \frac{J_1J_2}{N^{1+\delta}}\to \infty   \text{\ \ for some\ }\delta>0.
}
These limits extend the relations given in \refb{eq:box1} to the more general case.
We deduce that a large $N$ solution to the equation $G_3=1$ could only exist when the product $J_1J_2$ lies 
somewhere in the range
\bea
N < J_1 J_2 < N^{1+ \delta},
\eea 
for any small positive constant $\delta$.

By constraining $J_1J_2$ to lie within this range, the expression for the three-point correlator \refb{eq:3ptj1j2} 
can be simplified. Since we require $J_1J_2$ to be grow larger than $N$, and have constrained both $J_1$ and $J_2$ to be less than $N^{\frac{2}{3}}$, we must have that $J_1\gg N^\frac{1}{3}$, i.e. $J_1$ grows at least as quickly as a positive power of $N$. Also, the factors of the form $(1-e^{-x})$ in \refb{eq:3ptj1j2} tends to 1 if $x$ tends to $\infty$, so we can use the facts that $J_1J_2/N\to \infty$ near the threshold and $J_1 \leq J_2$ to neglect several factors and write
\bea
G_3 \sim \sqrt{\frac{J_1J_2}{(J_1+J_2)N}}\exp{\left(\frac{J_1J_2}{2N}\right)}\left(1-e^{-\frac{J_1^2}{N}}\right)^{-\frac{1}{2}}.
\eea
We can keep track of the errors generated in approximating the asymptotic form of the correlator by writing the \emph{exact} expression,
\bea
G_3 = \sqrt{\frac{J_1J_2}{(J_1+J_2)N}}\exp{\left(\frac{J_1J_2}{2N}\right)}\left(1-e^{-\frac{J_1^2}{N}}\right)^{-\frac{1}{2}} e^{-\frac{r}{2}},
\eea
where again the remainder function $r(J_1,J_2,N)$ is defined implicitly by this equation, and the $J_i$ scale with $N$ in the range $N^\frac{1}{3} \ll J_1 \leq J_2 \ll N^\frac{2}{3}$. This remainder function tends to zero with $N$,
but its leading-order behaviour will in general change depending on the scaling behaviour
of $J_1$ and $J_2$. We will later show that, near the threshold, the remainder function is of the order 
\bea
r = \cO\left(\frac{(\log N)^2}{J_1}\right),
\eea
and so decays to zero at a faster rate than some inverse power of $N$.

We wish to simplify the equation
\bea
G_3 = \sqrt{\frac{J_1J_2}{(J_1+J_2)N}}\exp{\left(\frac{J_1J_2}{2N}\right)}\left(1-e^{-\frac{J_1^2}{N}}\right)^{-\frac{1}{2}} e^{-\frac{r}{2}} = 1 \label{eq:3ptwithr}
\eea
in the large $N$ limit. A convenient way to do this is by using the Lambert $W$-function, and its large argument expansion. 
Equation \refb{eq:3ptwithr} is solved exactly (with the implicit remainder term $r$) by
\bea
\frac{J_1J_2}{N} = W\left((J_1+J_2)(1-e^{-\frac{J_1^2}{N}})e^{r}\right). \label{eq:wj1j2}
\eea
The argument of the $W$-function changes depending on the behaviour of 
$J_1^2/N$ with increasing $N$, but will grow to infinity in all relevant cases, allowing us to use the 
large argument asymptotic expansion of the $W$-function,
\bea
W(z) = \log z - \log \log z + \hot{\log \log z}{\log z}.
\eea
To proceed, we must consider three possible scaling behaviours of $J_1^2/N$ in turn: the case when $J_1^2/N$ tends to zero,
the case when $J_1^2/N$ tends to a constant, and the case when $J_1^2$ tends to infinity.

First, consider the case where $J_1^2/N\to 0$. We have
\bea
(1-e^{-\frac{J_1^2}{N}}) = \frac{J_1^2}{N} + \cO\left(\frac{J_1^2}{N}\right)^2
\eea
so
\bea
(J_1+J_2)(1-e^{-\frac{J_1^2}{N}}) = \frac{J_1J_2}{N}J_1\left( 1 + \frac{J_1}{J_2}\right)\left(1 + \hot{J_1^2}{N}\right)
\eea
which must tend to infinity since $J_1J_2/N$ and $J_1$ are large.
Neglecting the remainder term $r$ for the moment, we expand out the $W$-function to find the threshold equation
\begin{multline}
\frac{J_1J_2}{N} = \log J_1 - \log \log J_1 + \log \left(\frac{J_1J_2}{N}\right) \\
+ \log\left(1+ \frac{J_1}{J_2}\right) - \log \left[1 + \frac{1}{\log J_1}\left( \log \left(\frac{J_1J_2}{N}\right) + \log\left(1+ \frac{J_1}{J_2}\right)  \right) \right] \\ 
+ \hot{J_1^2}{N} + \hot{\log \log (J_1+J_2)(1-e^{-\frac{J_1^2}{N}}) }{\log (J_1+J_2)(1-e^{-\frac{J_1^2}{N}}) } \label{eq:j1small}
\end{multline}
This fairly involved expression can be substantially simplified as follows: first, we simplify the final error term by giving its leading behaviour
in terms of $N$. Next, we show that all terms on the second line are small at large $N$, which allows us to deduce that the leading-order 
behaviour of the expression is $\log J_1$. Finally, by plugging in $\log J_1(1+o(1))$ into the expressions for $J_1J_2/N$ on the RHS of
\refb{eq:j1small}, we will find that the $\log \log J_1$ term cancels, and that only one large term remains in its asymptotic series expansion.

First, we consider the latter remainder term. 
We know that $J_1$ and $J_2$ scale with $N$ at a larger rate than some positive power of $N$, so 
$\log J_1$ is $\cO\left(\log N\right)$ to leading order.
We've also required $J_1J_2/N$ to scale to infinity at a slower rate than any positive power of $N$, as this is required
for the threshold solution to $G_3=1$ to be valid at large $N$. 
This means that $\log (J_1J_2/N)$ must be $o(\log N)$. We deduce that
\bea
\log \left[(J_1+J_2)(1-e^{-\frac{J_1^2}{N}}) \right] &= & \log J_1 + \log\left(\frac{J_1J_2}{N}\right) + \log\left(1+ \frac{J_1}{J_2}\right) 
+ \hot{J_1^2}{N}
\\ &=& \cO(\log N),
\eea
and hence
\bea  
\hot{\log \log (J_1+J_2)(1-e^{-\frac{J_1^2}{N}}) }{\log (J_1+J_2)(1-e^{-\frac{J_1^2}{N}}) } = \hot{\log \log N}{\log N}.
\eea
Both this term and the $\hot{J_1^2}{N}$ term are small in the large $N$ limit.
Next, we can see that all terms on the second line of \refb{eq:j1small} must be small.
Noting that
\bea
\frac{J_1}{J_2} &=& \frac{J_1^2}{N}\frac{N}{J_1J_2} \to 0 \label{eq:j1j2ratio}
\eea
since $J_1^2/N\to 0$ and $J_1J_2/N \to \infty$, we have that 
\bea
\log\left(1+\frac{J_1}{J_2}\right) \to 0.
\eea
Also, it was required that $J_1J_2/N$ grows to infinity with $N$, but not as a positive power of $N$ or greater, so $\log(J_1J_2/N)=o(\log N)$. Since $\log J_1=\cO(\log N)$, this means that 
\bea
\frac{1}{\log J_1} \log\left( \frac{J_1J_2}{N}\right) \to 0,
\eea
and so the second term in the second line of \refb{eq:j1small} is also small.
The largest term in  \refb{eq:j1small} must therefore be $\log J_1$, which is of order $\cO(\log N)$.
Using this and \refb{eq:j1j2ratio}, we see that $J_1/J_2$ must be smaller than $\cO(1/\log N)$, 
and so we can collate all the remainders in the threshold expression into two terms; we find
\bea
\frac{J_1J_2}{N} = \log J_1 -\log \log J_1 + \log\left(\frac{J_1J_2}{N}\right) + \hot{\log \log N}{\log N} +  \hot{J_1^2}{N}.
\eea
By plugging in this expression for $J_1J_2/N$ into the third term, we can cancel the $\log\log J_1$ and obtain the leading-order threshold equation
\bea
\frac{J_1J_2}{N}= \log J_1 + \hot{\log \log N}{\log N} + \hot{J_1^2}{N}.
\eea
This formula is valid at the threshold, provided that $J_1^2/N\to 0$ with large $N$.
There are two different remainder terms in this expression as we have not imposed enough conditions on $J_1$ to state which term is larger.
Constraining the scaling behaviour of $J_1$ with $N$ would allow us to deduce which term is subleading.
For example, if we set $J_1 \sim \sqrt{N} (\log \log N)/ \sqrt{ \log N}$, then the $\cO(J_1^2/N)$ term is the leading error, but if 
$J_1 \sim N^{5/12}$ then the $\hot{\log \log N}{\log N}$ term is the largest error.

Next, we consider the case where $J_1^2/N$ tends to a constant. Starting from threshold equation
\bea
\frac{J_1J_2}{N} = W((J_1+J_2)(1-e^{-\frac{J_1^2}{N}})e^{r}),
\eea
the argument of the $W$-function is clearly large since $J_1+J_2$ grows with $N$.
Again neglecting the remainder term, we can use the large argument expansion of the $W$-function
and write
\begin{multline}
\frac{J_1J_2}{N} \sim \log \left((J_1+J_2)(1- e^{-\frac{J_1^2}{N}})\right) - \log\log \left((J_1+J_2)(1- e^{-\frac{J_1^2}{N}})\right) \\
+ \hot{\log \log \left((J_1+J_2)(1- e^{-\frac{J_1^2}{N}})\right) }{\log \left((J_1+J_2)(1- e^{-\frac{J_1^2}{N}})\right) }
\end{multline}
Since $\log((J_1+J_2)(1-e^{-\frac{J_1^2}{N}})) = \cO(\log N)$, we can simplify this remainder term and expand out the second term to write 
\bea
\frac{J_1J_2}{N} \sim \log(J_1+J_2) - \log \log (J_1+J_2) + \log\left(1-e^{-\frac{J_1^2}{N}}\right) + \hot{\log \log N}{\log N}. \label{eq:j1plusj2}
\eea
In writing this expression, we have dropped a term of $\hot{1}{\log N}$ as it is subleading to the $\hot{\log \log N}{\log N}$ remainder term.
The first two terms in this expression grow large with increasing $N$, and the third term tends to a constant. 

Finally, we consider the case where $J_1^2/N$ tends to infinity with $N$. Again we find that 
\refb{eq:j1plusj2} still holds, but that the third term now tends to zero. From the series expansion of the 
logarithm, we have
\bea
\log\left(1-e^{-\frac{J_1^2}{N}}\right) = \cO\left(e^{-\frac{J_1^2}{N}}\right),
\eea
so we write the final expression 
\bea
\frac{J_1J_2}{N} \sim \log (J_1+J_2) - \log \log (J_1+J_2) + \cO\left( e^{-\frac{J_1^2}{N}} \right) + \hot{\log \log N}{\log N}.
\eea
Again, we have two remainder terms, as we have not specified how quickly $J_1^2/N$ scales to infinity with $N$
and so cannot state which is the larger.

Summarizing the above, we have three different threshold equations for the different regimes of $J_1^2/N$.
Listed in order of increasing $J_1^2/N$, we have:
\bea
\frac{ J_1 J_2}{N} = \left\{ \begin{array}{ll}
\log J_1 + \hot{\log \log N}{\log N} + \hot{J_1^2}{N} & \frac{J_1^2}{N}\to 0 \\
\log (J_1+J_2)  -\log \log (J_1+J_2) + \log\left(1-e^{-\frac{J_1^2}{N}}\right) + \hot{\log \log N}{\log N} & \frac{J_1^2}{N} \to \mbox{const.} \\
\log (J_1+J_2) - \log \log (J_1+J_2) + \hot{\log \log N}{\log N} + \cO\left(e^{-\frac{J_1^2}{N}}\right) & \frac{J_1^2}{N} \to \infty. \\ \end{array} \right. \label{eq:solnarray}
\eea
In all cases, the explicitly-given terms are non-zero in the large $N$ limit, and the higher-order terms are small. 
All these large terms are necessary to describe the threshold accurately at large $N$;
if we plug \refb{eq:solnarray} into \refb{eq:3ptwithr} with the remainder terms and $r$ discarded, 
then the correlator tends to one at large $N$ in each case.

The angular momenta $J_1$ and $J_2$ grow at least as quickly as a positive power of $N$, so the leading-order
term in the threshold is always proportional to $\log N$. 
If we assume that the power-dependence of $J_1$ on $N$ is simple enough that it can be separated out into the form
$J_1= N^{\alpha_1}e^{\delta_1}$, where $\alpha_1$ is a constant and $|\delta_1(N)| \ll \log N$, 
then the leading-order term of the threshold solution is
\bea
J_1 J_2 \sim \alpha_1 N\log N.
\eea

We have so far neglected the error parameter $r$ without discussion, but we can now justify this. 
To derive the equation
\bea
G_3 \sim \sqrt{\frac{J_1J_2}{(J_1+J_2)N}}\exp\left(\frac{J_1J_2}{2N}\right) \left(1-e^{-\frac{J_1^2}{N}}\right)^{-\frac{1}{2}}
\eea
near the threshold, we have dropped corrections of at most order $\cO\left(1/J_1\right)$ and $\cO(J_1J_2(J_1+J_2)/N^2)$. 
Near the threshold, $J_1$ and $J_2$ satisfy
\bea
\frac{J_1J_2(J_1+J_2)}{N^2} \sim \frac{1}{J_1}\left(\frac{J_1J_2}{N}\right)^2 = \cO\left(\frac{(\log N)^2}{J_1}\right).
\eea
The remainder parameter $r$, defined in \refb{eq:3ptwithr}, must contain all the corrections to the correlator 
near the threshold. We can therefore state that, near the threshold, the largest corrections to $r$ must be 
\bea
r = \hot{(\log N)^2}{J_1},
\eea
which decays to zero with $N$ at a faster rate than some inverse power of $N$.
If we reintroduce this remainder when expanding out the $W$-function in \refb{eq:wj1j2}, we will
modify each equation in \refb{eq:solnarray} by the addition of an $r$ term, plus $\cO(r^2)$ corrections.
However, this term must be smaller than $\hot{\log \log N}{\log N}$, and in fact is smaller than any
power of $(\log \log N / \log N)$: in terms of the parameter $v=\log N$, the contributions from $r$ are
exponentially suppressed in $v$. As a consequence, we can always drop these terms
from the solution.

\subsection{A change of variables}

The threshold equation $G_3(J_1,J_2,N)=1$ defines a two-dimensional threshold 
surface in three-dimensional $(J_1,J_2,N)$-space. We can develop some insight into 
the relation between this surface and the physical properties of the correlator by changing 
the parameter space variables.
If we take $N$ to be fixed but large enough that the remainder $\hot{\log \log N}{\log N}$ is small, then we can 
use \refb{eq:solnarray} to rewrite the threshold as a curve in $E=J_1+J_2$ and $\Delta J=|J_2-J_1|$. 
For the region where $J_1^2/N\to 0$, i.e. $(E^2 - \Delta J^2)/N  \to 0$, then the threshold of factorization is
\bea
\frac{E^2 - \Delta J^2}{4N\log\left(E-\Delta J\right) } \sim 1 + \hot{\log \log N}{\log N},
\eea
and for the region where $J_1^2/N = (E^2 - \Delta J^2)/N$ does not tend to zero, then the threshold is
\bea
\frac{E^2-\Delta J^2}{4N(\log E - \log \log E) } \sim 1+ \hot{\log \log N}{\log N},
\eea
where all the discarded terms are small. 

We can say something about how perturbations away from the threshold in $(E, \Delta J, N)$ space affect the factorization of the correlator by rewriting the correlator in the form
\bea
G_3(E, \Delta J, N) = \left[\frac{E^2-\Delta J^2}{4NE}\exp\left( \frac{E^2 - \Delta J^2}{4N}\right)\left(1-e^{-\frac{1}{4N}(E-\Delta J)^2} \right)^{-1}e^{-r}\right]^\frac{1}{2}.
\eea
It is convenient to work with $\log (G_3)^2$, and allow $E$ and $\Delta J$ to be independent of $N$. Taking the differential of 
$\log (G_3)^2$, we have
\bea
d\log(G_3)^2 = \frac{2}{G_3}d G_3= \frac{2}{G_3}\left[\frac{\partial G_3}{\partial E}dE +\frac{\partial G_3}{\partial (\Delta J)}d(\Delta J) +\frac{\partial G_3}{\partial N}dN\right],
\eea
Expressing the coefficients of the differentials in terms of $J_1, J_2$ and $N$ for convenience, we have
\begin{multline}
d\log (G_3)^2 = \frac{1}{2}dE \left[ \frac{J_2}{N} + \frac{1}{J_1} + \frac{J_1}{N} + \frac{1}{J_2} - \frac{2}{J_1+J_2} - \frac{2J_1}{N(e^{\frac{J_1^2}{N}}-1)} \right] \\
+ \frac{1}{2}d(\Delta J)\left[ -\frac{J_2}{N} - \frac{1}{J_1} + \frac{J_1}{N} + \frac{1}{J_2} + \frac{2J_1}{N(e^{\frac{J_1^2}{N}}-1)}\right] \\
- \frac{dN}{N}\left[ \frac{J_1J_2}{N} + 1 - \frac{J_1^2}{N(e^{\frac{J_1^2}{N}}-1)}  \right] - dr.
\end{multline}
At large $N$ and near the threshold $J_1J_2 = \cO(N \log N)$, the largest term in the coefficient of $dE$ is $J_2/N$, which is of order $\hot{\log N}{J_1}$. This means that $\frac{\partial G_3}{\partial E}$ is positive at large $N$. Similarly, the largest term in the coefficient of $d(\Delta J)$ is $-J_2/N$, which is order $\hot{\log N}{J_1}$, and so $\frac{\partial G_3}{\partial \Delta J}$ is negative at large $N$. The corrections to $dE$ and $d(\Delta J)$ from the differential of the error function $dr$ are order $\hot{\log N}{J_1^2}$ at the threshold, and so are subleading.

The signs of the partial derivates of $G_3$ with respect to $E$ and $\Delta J$ gives us some interesting insights into factorization near the threshold.
If we consider $N$ to be large and fixed, and take $E$ and $\Delta J$ near to the threshold, then a small increase in the energy of the single graviton $E$ will increase the correlator $G_3$, and move the correlator into the breakdown region. On the other hand, if the separation between the gravitons $\Delta J$ in the multi-graviton state is increased by a small amount, then $G_3$ will decrease, and the correlator will move into the factorization region.

\section{Non-extremal correlators}\label{sec:next}
We can consider the existence of a threshold of factorization for a non-extremal three-point function with operators formed from 
the complex scalar fields $Z=\phi_5+i\phi_6$ accompanied by a small number of $Y=\phi_3+i\phi_4$ insertions. 
Consider a correlator of symmetrized trace operators inserted at the points $x_1$, $x_2$, and $y$:
\bea
\cor{\str(Z^{J_1}Y^{J_3})(x_1)\str(Z^{J_2}Y^{\dagger J_3})(x_2) \tr(Z^{\dagger J_1+J_2})(y)}.
\eea
In a similar manner to the extremal correlator consisting of only $Z$-fields, we can
use the conformal symmetry to separate out a position-independent correlator 
by a particular choice of operator insertion locations.
Under the inversion $y\to y'=y/|y|^2$, the antiholomorphic operator transforms as
\bea
\tr(Z^{\dagger J_1+J_2})(y) &\to& \tr'(Z^{\dagger J_1+J_2})(y') \ret &=&
|y|^{J_1+J_2}\tr(Z^{\dagger J_1+J_2})(y).
\eea
By taking $x_1\to 0$ and $y'\to 0$ i.e. $y\to \infty$, the correlator becomes
\bea
\cor{\str(Z^{J_1}Y^{J_3})(0)\str(Z^{J_2}Y^{\dagger J_3})(x_2) \tr'(Z^{\dagger J_1+J_2})(0)} \nn \\ = \frac{
 \cor{\str(Z^{J_1}Y^{J_3})\str(Z^{J_2}Y^{\dagger J_3})\tr(Z^{\dagger J_1+J_2})} 
 }{|x_2|^{2J_3} }.
\eea
We have separated out a combinatoric factor which can be evaluated by a matrix model calculation.
Unlike the extremal correlator, however, the separation $|x_2|$ between the operators inserted at $\str(Z^{J_1}Y^{J_3})$ and $\str(Z^{J_2}Y^{\dagger J_3})$ is still present in this correlator.
Introducing the notation $\epsilon\equiv |x_2$ for the magnitude of the separation between these two operators,
and $\parallel \cO \parallel = \sqrt{\cor{\cO \cO^\dagger}}$ for the norm of a matrix model
operator $\cO$, then the multiparticle-normalized correlator is
\bea
G_3(J_i, N; \epsilon) = \frac{ \cor{\str(Z^{J_1}Y^{J_3})\str(Z^{J_2}Y^{\dagger J_3})\tr(Z^{\dagger J_1+J_2})} }
{\epsilon^{2J_3}  \norm{\str(Z^{J_1}Y^{J_3})} \norm{\str(Z^{J_2}Y^{J_3})} \norm{\tr(Z^{J_1+J_2})} }
\eea
The appearance of this position-dependence means that the three-point correlator is dimensionful, and so it is not meaningful
to define the threshold as being when the correlator approaches a fixed number at large $N$. 
However, if we introduce an arbitrary mass scale $\Lambda$, then we can instead consider the
combination $\Lambda^{-2J_3} G_3(J_i, N; \epsilon)$, which is dimensionless.
We define the \emph{non-extremal threshold} as the solution to the equation
\bea
\Lambda^{-2J_3}G_3(J_i, N; \epsilon) = 1.
\eea 

A natural choice of $\Lambda$ would be a UV cutoff of the CFT. This
will modify correlators in general, and the $\epsilon^{-2J_3}$ factor will be modified to 
\bea 
\frac{ 1 }{ \epsilon^{2J_3} }\left(1 + o(\epsilon^{-1} \Lambda^{-1}) \right).
\eea
The higher-order corrections can be neglected if we require that the separation $\epsilon$ is much larger than 
the cutoff length $\Lambda^{-1}$.
We can do this by setting $\epsilon \Lambda$ to be large and independent of $N$, or by allowing $\epsilon \Lambda$ to grow large
with $N$. 
It is convenient in the following to define $\cR:= \epsilon\Lambda$ as the dimensionless ratio between the cutoff separation
and the length scale.
This is required to be large for the higher-order corrections to $\epsilon$ to be absent.
The non-extremal threshold equation can then be written in the form
\bea
\Lambda^{-2J_3}G_3(J_i, N; \epsilon) = \cR^{-2J_3}  \frac{ \cor{\str(Z^{J_1}Y^{J_3})\str(Z^{J_2}Y^{\dagger J_3})\tr(Z^{\dagger J_1+J_2})} }
{\norm{\str(Z^{J_1}Y^{J_3})} \norm{\str(Z^{J_2}Y^{J_3})} \norm{\tr(Z^{J_1+J_2})} } = 1. \label{eq:nextthreshold}
\eea

To investigate the threshold of this non-extremal correlator, we look for an exact finite 
$N$ expression of the correlator that is valid when some of the operator dimensions are large. 
There are three matrix model correlator expressions that we need in order to evaluate the correlator:
\bea
\norm{\str(Z^{J_1}Y^{J_3})}^2\ = \cor{\str(Z^{J_1}Y^{J_3}) \str(Z^{\dagger J_1}Y^{\dagger J_3})}, \ret
\norm{\tr(Z^{J_1+J_2})}^2\ = \cor{\tr(Z^{J_1+J_2})\tr(Z^{\dagger J_1+J_2})}, \ret
\cor{\str(Z^{J_1}Y^{J_3}) \str(Z^{J_2}Y^{\dagger J_3})\tr(Z^{\dagger J_1+J_2})}.\quad 
\eea
The norm $\norm{\tr(Z^{J_1+J_2})}^2$ is known explicitly, but we have not found a closed form of the other correlators 
for general operator dimensions. However, exact evaluations of the correlator can be found for small values of $J_3$, 
where there is only a small number of $Y$-insertions; in the following we focus on the `near-extremal' case when $J_3=1$.

\subsection{The `near-extremal' correlator}
We set $J_3=1$ in \refb{eq:nextthreshold} and consider the correlator 
\bea
G_3(J_i, N; \epsilon) = \frac{ \cor{\str(Z^{J_1}Y)\str(Z^{J_2}Y^{\dagger })\tr(Z^{\dagger J_1+J_2})} }
{\epsilon^{2}  \norm{\str(Z^{J_1}Y)} \norm{\str(Z^{J_2}Y)} \norm{\tr(Z^{J_1+J_2})} }.\label{eq:nonext}
\eea
The norm $\norm{\tr(Z^{J_1+J_2})}^2$ was known previously \cite{kpss} and used in Sections \ref{sec:j1eqj2} and \ref{sec:j1neqj2}:
\bea
\cor{\tr(Z^{J_1+J_2})\tr(Z^{\dagger J_1+J_2})} = (J_1+J_2)!\left[ \binom{N+J_1+J_2}{J_1+J_2+1} - \binom{N}{J_1+J_2+1}\right].
\eea
For $J_3=1$, there is only one pair of $Y$-matrices, so the contraction of the three-point function can be performed 
immediately. The unnormalized three-point correlator becomes
\bea
\contraction{ \langle \str(Z^{J_1} }{Y}{)  \str(Z^{J_2} }{{Y}^{\dagger}}
\cor{ \str(Z^{J_1} Y)  \str(Z^{J_2} {Y}^{\dagger})  \tr(Z^{\dagger J_1+J_2}) } = \cor{\tr (Z^{J_1+J_2} )
\tr (Z^{\dagger J_1+J_2} ) },
\eea
where we have used the fact that $\str(Z^{J_1+J_2}) = \tr(Z^{J_1+J_2} )$. This means that \refb{eq:nonext} reduces to
\bea
G_3(J_i, N; \epsilon) = \frac{  \norm{\tr(Z^{J_1+J_2})}  }
{\epsilon^{2}  \norm{\str(Z^{J_1}Y)} \norm{\str(Z^{J_2}Y)} }. \label{eq:nonext2}
\eea
The other correlators can be determined by tensor space methods. In Appendix \ref{sec:appmm},
we have derived the equation
\bea
\cor{ \str(Z^{J_1} Y^{J_2})  \str(Z^{\dagger J_1} Y^{\dagger J_2})  }\  = J_1!J_2! \left[ \binom{N+J_1+J_2}{ J_1+J_2+1} - \binom{N}{ J_1+J_2+1} \right].
\eea
Substituting in the relevant values of $J_1$ and $J_2$ in to the correlators in the denominators of \refb{eq:nonext}, we
find that 
\bea
\norm{ \str(Z^{J_1} Y) } &=& \sqrt{J_1!} \left[ \binom{N+J_1+1}{J_1+2} - \binom{N}{J_1+2} \right]^\frac{1}{2}, \\ 
\norm{ \str(Z^{J_2} Y) } &=& \sqrt{J_2!} \left[ \binom{N+J_2+1}{J_2+2} - \binom{N}{J_2+2} \right]^\frac{1}{2},
\eea
and so
\ml{
\Lambda^{-2} G_3(J_i, N; \epsilon)
= \cR^{-2} \left(\frac{(J_1+J_2)!}{J_1!J_2!}\right)^\frac{1}{2}\left[ \binom{N+J_1+J_2}{J_1+J_2+1} - \binom{N}{J_1+J_2+1}\right]^\frac{1}{2} \times
\\ \times \left[ \binom{N+J_1+1 }{ J_1+2} - \binom{N}{J_1+2}\right]^{-\frac{1}{2}}\left[ \binom{N+J_2+1}{ J_2+2} - \binom{N}{J_2+2}\right]^{-\frac{1}{2}}.
}

This is the finite $N$ expression of the non-extremal correlator when $J_3=1$. It is valid for small or large $J_1$ and $J_2$, 
provided that $J_1, J_2 \ll N$. 
As in the extremal case, we wish to find the asymptotic form of this expression when $J_1$, $J_2$, and $N$ are large, but the ratios $J_1/N$ and $J_2/N$ are small.
Making the assumptions that $J_1\leq J_2 \ll N^{\frac{2}{3}}$, then equation \refb{eq:infsumdrop} still holds 
with $J$ replaced by $J_1+1$. Dropping the subleading corrections, we find that
\bea
J_1!\left[\binom{N+J_1+1}{J_1+2} - \binom{N}{J_1+2}\right] \sim \frac{N^{J_1+2}}{J_1^2}\exp\left(\frac{J_1^2}{2N}\right)\left(1 - e^{-\frac{J_1^2}{N}}\right),
\eea
and similarly for $J_2$. The full large $N$ expression for the correlator for $1\ll J_1\leq J_2 \ll N$ is therefore
\bea
\Lambda^{-2} G_3 \sim  \cR^{-2}\sqrt{\frac{J_1^2J_2^2}{(J_1+J_2 )N^3}}\exp\left(\frac{J_1J_2}{2N}\right)
\sqrt{\frac{\left(1 - e^{-\frac{(J_1+J_2)^2}{N}}\right)  }{\left(1 - e^{-\frac{J_1^2}{N}}\right) \left(1 - e^{-\frac{J_2^2}{N}}\right) }}
\eea

We can argue that the correlator must decay to zero if $J_1J_2/N$ is small as follows: If $J_1J_2/N$ tends to zero with $N$, then
the exponential term tends to 1. The factor $\cR^{-2}$ has already been taken to be small.
Since $J_1J_2/N$ is small and we have assumed that $J_1\leq J_2$, we know that $J_1^2/N$ is also small and so
\bea 1-e^{-\frac{(J_1+J_2)^2}{N}} &\sim& 1-e^{-\frac{J_2^2}{N}}, \ret
\frac{J_1^2/N}{1-e^{-\frac{J_1^2}{N}}} &\sim& 1,
\eea
and thus we can deduce that
\bea
\Lambda^{-2}G_3 \sim \cR^{-2}\frac{J_2}{N}\frac{1}{\sqrt{J_1+J_2}}.
\eea
The correlator must therefore tend to zero when $J_1J_2/N$ is small.

On the other hand, consider the case when $J_1J_2/N$ grows larger than a positive power of $N$, i.e. $J_1J_2 > N^{1+\delta}$ for some $\delta>0$. The exponential term will dominate the expression, as it will grow to infinity
exponentially quickly with $N$ as compared to the other factors of $J_1, J_2$, and $N$ outside of the exponential. In this case, the correlator must definitely grow to infinity (provided that $\cR$ is does not grow with $N$ at a faster than a power of $N$).
Summarizing the above, we have 
\boxeq{
G_3(J_1,J_2, N; \cR) \to 0,& \qquad \frac{J_1J_2}{N}\to 0, \ret
G_3(J_1,J_2, N; \cR) \to \infty,& \qquad \frac{J_1J_2}{N^{1+\delta}}\to \infty   \text{\ \ for some\ }\delta>0.
}
The threshold must therefore be constrained to lie in the region 
\bea
N < J_1J_2 < N^{1+\delta}.
\eea
In this range, the large $N$ behaviour of the correlator is simply
\bea
\Lambda^{-2} G_3(J_1,J_2, N; \epsilon) \sim \cR^{-2}\sqrt{\frac{J_1^2J_2^2}{(J_1+J_2)N^3}}\exp\left(\frac{J_1J_2}{2N}\right)
\left(1 - e^{-\frac{J_1^2}{N}}\right)^{-\frac{1}{2}}.
\eea

We can encompass all the errors present in approximating this expression by the function $r$, defined by the equation
\bea
\Lambda^{-2}G_3(J_1,J_2, N; \epsilon) = \cR^{-2}\sqrt{\frac{J_1^2J_2^2}{(J_1+J_2)N^3}}\exp\left(\frac{J_1J_2}{2N}\right)
\left(1 - e^{-\frac{J_1^2}{N}}\right)^{-\frac{1}{2}}e^{-\frac{r}{2}}, \label{eq:nonext3}
\eea
and attempt to solve asymptotically the threshold equation
\bea
\Lambda^{-2}G_3(J_1,J_2, N; \epsilon) = 1.
\eea
We consider the cases $J_1=J_2$ and $J_1\neq J_2$ separately.

\subsection{\texorpdfstring{$J_1=J_2$}{J1=J2}}
If we consider the non-extremal correlator when $J_1=J_2=J$, then the threshold equation with error function $r$ becomes 
\bea
\cR^{-2}\sqrt{\frac{J^3}{2N^3}}\exp\left(\frac{J^2}{2N}\right)
\left(1 - e^{-\frac{J^2}{N}}\right)^{-\frac{1}{2}}e^{-\frac{r}{2}} = 1.
\eea
This has an exact solution in term of the $W$-function,
\bea
\frac{J_t^2}{N} = \frac{3}{2}W\left[\frac{2^{5/3}}{3}N \cR^{8/3}\left(1-e^{-\frac{J_t^2}{N}}\right)^\frac{2}{3}e^{\frac{2r}{3}}\right].
\eea
The argument of the $W$-function must be large, so we can again expand it in terms of logarithms.
The factors of $(1-e^{-\frac{J^2}{N}})$ and $e^\frac{2r}{3}$ must be subleading, and so a short calculation
shows that the threshold expands out to
\bea
\frac{J_t^2}{N} = \frac{3}{2}\log N + 4\log \cR - \frac{3}{2}\log \left[\log N + \frac{8}{3}\log \cR\right] + \frac{1}{2}\log\left(\frac{32}{27}\right)
+ o(1).
\eea

In this large $N$ expansion of the threshold, we have the two parameters $N$ and $\cR\equiv \epsilon \Lambda$. If we take $\cR$ to be large
but independent of $N$, then it must become subleading in the large $N$ limit, and the threshold becomes
\bea
\frac{J_t^2}{N} &=& \frac{3}{2}\log N - \frac{3}{2}\log \log N + 4\log \cR + \frac{1}{2}\log\left(\frac{32}{27}\right) + o(1) \label{eq:logr} \\
&\sim & \frac{3}{2}\log N.
\eea
Alternatively, we can allow the ratio $\cR$ to grow large with $N$, by letting either the separation of the operators $\epsilon$
or the cutoff scale $\Lambda$ grow with $N$. The $\log \cR$ terms are subleading and the above expression still
holds if $\cR$ scales to infinity at a slower rate than a power of $N$. If $\cR$ grows like a power of $N$, then it can influence the 
leading constant of the threshold, but it is still logarithmically dependent on $N$. In all these cases,
the leading-order behaviour of the threshold is simply
\bea
J^2 = \cO(N \log N),
\eea
as was the case for the extremal correlator.

The expansion of the threshold given in \refb{eq:logr} tells us something new about the factorization thresholds for non-extremal
correlators. 
The $(4\log \cR)$ term, which did not appear in the extremal threshold, means that the threshold in the non-extremal
case depends on the separation of the correlators in the boundary directions.
If we considered a system at the threshold at fixed large $N$ and fixed large $\cR$, then a decrease in $\cR$
will lead to an increase in $\Lambda^{-2}G_3$, and an increase in $\cR$ will lead to a decrease in $\Lambda^{-2}G_3$.
From the bulk AdS perspective, this means that we move from factorization to breakdown
when the gravitons are moved closer together in the boundary directions, perpendicular to the AdS radius.

\subsection{\texorpdfstring{$J_1 \neq J_2$}{J1=/= J2}}
When $J_1$ and $J_2$ are not equal, but lie in the region $N^\frac{1}{3} \ll J_1 \leq J_2 \ll N^\frac{2}{3}$, then equation \refb{eq:nonext3} has the solution
\bea
\frac{J_1J_2}{N} = 2W\left[\sqrt{\frac{1}{4}\cR^4 N(J_1+J_2)(1-e^{-\frac{J_1^2}{N}})e^r}\right].
\eea
The form of the expansion of the $W$-function depends on the scaling behaviour of the smallest
angular momentum with $N$, which we have chosen to be $J_1$. We consider separately three cases:
$J_1^2/N$ tends to zero, $J_1^2/N$ tends to a constant, and $J_1^2/N$ tends to infinity.

If $J_1^2/N\to 0$, then the leading terms in the expansion of the $W$-function are 
\bea
\frac{J_1J_2}{2N} &\sim& \frac{1}{2}\log\left[\frac{1}{4}\cR^4 N (J_1+J_2)\frac{J_1^2}{N}\right] 
- \log \left[\frac{1}{2}\log\left[\frac{1}{4}\cR^4 N (J_1+J_2)\frac{J_1^2}{N}\right]\right] + o(1)
\eea
\bea
= \frac{1}{2}\log\left(\frac{J_1N\cR^4}{4}\right) + \log\left(1+\frac{J_1}{J_2}\right) + \log\left(\frac{J_1J_2}{N}\right)
- \log \left[ \frac{1}{2}\log\left[\frac{1}{4}\cR^4 N (J_1+J_2)\frac{J_1^2}{N}\right]\right]
\eea
Plugging in $J_1J_2/N$ into the third term, the log-log cancels and we have
\bea
\frac{J_1J_2}{2N} = \frac{1}{2}\log\left(\frac{J_1N\cR^4}{4}\right) + \log\left(1+\frac{J_1}{J_2}\right) + \cO(1).
\eea
Since $J_1\leq J_2$, the second term is $\cO(1)$, hence
\bea
\frac{J_1J_2}{N} = \log (J_1N) + 4\log \cR +\cO(1).
\eea

If $J_1^2/N$ tends to a constant at large $N$, then the expansion becomes
\bea
\frac{J_1J_2}{2N} &=& \frac{1}{2}\log\left[\frac{\cR^4N}{4}(J_1+J_2)\right] + c
- \log\left[\frac{1}{2}\log\left[\frac{\cR^4N}{4}(J_1+J_2)\right] + c \right] + o(1),
\eea
where $c$ is some constant (order 1 with respect to $N$). 
Hence
\bea
\frac{J_1J_2}{N} = \log ((J_1+J_2) N) + 4\log \cR - \log \log ((J_1+J_2) N ) + \cO(1).
\eea
If $J_1^2/N$ tends to infinity with $N$, then the above equation also holds but with $c$ replaced by zero.

We can collate these three cases into a single equation by taking the leading scaling-behaviour of $J_1$ to be fixed, i.e. assuming $J_1=N^{\alpha_1}e^{\delta_1}$ for subleading $\delta_1$ and constant $\alpha_1$.
The threshold can then be written in all cases as 
\bea
\frac{J_1J_2}{N} = (1+\alpha_1)\log N + 4 \log \cR + o(\log N).
\eea

As in the extremal case, decreasing the difference between the angular momenta $\Delta J=|J_2-J_1|$ 
will move the correlator from the threshold to the breakdown region.
In addition, from the structure of the correlator in \refb{eq:nonext3}, it is clear that
decreasing $\cR$ while fixing $N$, $J_1$, and $J_2$ will move the correlator from the threshold to the breakdown region. 
From the bulk AdS point of view, non-extremal correlators correspond to the interactions of 
Kaluza-Klein gravitons with angular momenta in perpendicular directions in the $S^5$.
We can move from the threshold to the breakdown region by moving the gravitons closer together in the boundary directions, or by decreasing the separation in the graviton energies.

\section{Multi-gravitons and non-trivial  backgrounds }\label{sec:future}
In the previous sections we have studied in detail the thresholds of some simple extremal and non-extremal three-point functions.
In this section we briefly discuss two other examples of extremal correlators for which we have found explicit expressions of the threshold: a correlator corresponding to a $k+1$-graviton system, and a correlator corresponding to gravitons in an LLM background. 
We find a very similar form of the thresholds to the previous examples in both cases.
In the future, developing the tools to calculate more general correlators in the half-BPS sector could give us more insight
into general properties of thresholds, and thus also shed light on the behaviour of high-momentum graviton systems in supergravity.

\subsection{The \texorpdfstring{$k+1$}{k+1}-graviton correlator}
We can calculate the extremal correlator associated to $k$ gravitons scattering into a single graviton,
\bea 
\corb{\prod_{i=1}^k (\tr Z^{J_i}) \tr Z^{\dagger \sum J_i}}
\eea
and take the large dimensions limit using similar techniques.
An outline of the derivation of the $k\to 1$ correlation function and its large $N$ limit is given in Appendix \ref{sec:appmm}.
In the regime where all $J_i\ll N^\frac{2}{3}$  for all $i=1, 2,\ldots k$, then the correlator is asymptotic to
\bea
\sqrt{\frac{J_1\ldots J_k}{N^{k-1} \sum_i J_i}} \frac{(1-e^{-\frac{J_1\sum J_i}{N}})\ldots (1-e^{-\frac{J_k\sum J_i}{N}})}{ \sqrt{(1-e^{-\frac{J_1^2}{N}})\ldots (1-e^{-\frac{J_k^2}{N}})(1-e^{-\frac{(\sum J_i)^2}{N}})} } \exp\left( \sum_{i<j} \frac{J_i J_j}{2N}\right).
\eea
The factors in front of the exponential tend to zero as a power of $N$ when $1\ll J_i\ll N^\frac{2}{3}$. If \emph{all} pairs of dimensions satisfy $J_iJ_j\lesssim N$, then the exponential term is small, and the correlator decays to zero. However, if \emph{any} pair of distinct dimensions 
satisfy $J_iJ_j \geq N^{1+\delta}$ for some $\delta>0$, then the exponential term dominates any power of $N$, and so the correlator
tends to infinity. We can deduce that the factorization threshold when $G_3=1$ should be located when the product of the largest two
operators grows logarithmically larger than $N$:
\bea
J_iJ_j = \cO(N\log N).
\eea

In the case when all the $J_i$ are taken to be equal to $J$, then we can solve the threshold explictly at leading order. The correlator for $N^\frac{1}{2}< J < N^{\frac{1}{2}+\delta}$ is asymptotically
\bea
G_{k+1} \sim \sqrt{\frac{J^{k-1}}{kN^{k-1}}}\exp\left(\frac{k(k-1)}{4}\frac{J^2}{N}\right),
\eea
and the leading-order terms in the expansion of the threshold satisfying $G_{k+1}(J_t(N), N)=\nolinebreak1$ are
\bea
J_t^2 = \frac{N}{k}\left[\log N - \log \log N + \left(\frac{k+1}{k-1}\right)\log k + o(1)\right].
\eea
This can be interpreted as saying that as the number of gravitons \emph{increases}, the region in which factorization holds \emph{shrinks}.
When more gravitons are added to a system, they will start behaving like a single particle located further away from the boundary.

\subsection{Factorization thresholds for large backgrounds}

Thresholds of factorization can be considered in more general half-BPS bulk backgrounds,
specified in the dual description by Schur Polynomials. For a background described by a Young tableau 
$B$ with $n$ boxes, the associated Schur polynomial $\chi_{B}$ is a $U(N)$ character \cite{cjr},
\bea
\chi_{B}(Z) = \frac{1}{n!}\sum_{\sigma \in S_{n}}\chi_{B}(\sigma)\tr(\sigma Z^{\otimes n}).
\eea 
The CFT state corresponding to such a background is $\ket{B} = \chi_{B}(Z^\dagger)\ket{0}$, 
and the operator in this background are defined by \cite{dmk09}
\bea
\cor{\cO\ldots \cO}_B = \frac{\bra{B} \cO\ldots \cO \ket{B} }{\braket{B}{B}}.
\eea
This gives us the definition of a three-particle normalized correlator in the state,
\ml{
G_3(J_1,J_2; N,M)_B = \frac{ \cor{\tr Z^{J_1}\tr Z^{J_2}\tr Z^{\dagger J_1+J_2}}_B }{\sqrt{\cor{\tr Z^{J_1}\tr Z^{\dagger J_1}}_B \cor{\tr Z^{J_2}\tr Z^{\dagger J_2}}_B \cor{\tr Z^{J_1+J_2}\tr Z^{\dagger J_1+J_2}}_B}} \\
= \sqrt{\braket{B}{B}} \frac{ \bra{B} \tr Z^{J_1}\tr Z^{J_2}\tr Z^{\dagger J_1+J_2} \ket{B} }{ \sqrt{ \bra{B} \tr Z^{J_1}\tr Z^{\dagger J_1} \ket{B}\bra{B} \tr Z^{J_2}\tr Z^{\dagger J_2} \ket{B}\bra{B} \tr Z^{J_1+J_2}\tr Z^{\dagger J_1+J_2} \ket{B} } }. \label{newnorm}
}
One of the easiest ones backgrounds in which to perform the threshold calculation is the background corresponding to a large rectangular 
Young diagram with $N$ rows of length $M$, where $M$ is of the same order as $N$. 
In \cite{dmk09}, it was shown by performing manipulations of Schurs that the large rectangular background modifies the normalized correlator by shifting the matrix rank parameter from $N$ to $M+N$. That is, we have 
\bea
G_3(J_1,J_2; N,M)_B &=& \left[\frac{ \cor{\tr Z^{J_1}\tr Z^{J_2}\tr Z^{\dagger J_1+J_2}} }{\sqrt{\cor{\tr Z^{J_1}\tr Z^{\dagger J_1}} \cor{\tr Z^{J_2}\tr Z^{\dagger J_2}} \cor{\tr Z^{J_1+J_2}\tr Z^{\dagger J_1+J_2}}}}\right]_{N\rightarrow N+M} \\
&=& G_3(J_1,J_2, N+M).
\eea
Hence, the correlator in a large rectangular background only differs from the normalized correlator in that the argument $N$ is replaced by $N+M$. This means that, in this background, the threshold of factorization is at
\bea
J_1J_2 \approx (N+M) \log (N+M).
\eea
We interpret this as evidence that the presence of a background can \emph{increase} the size of the region in which factorization is valid.

\section{Conclusions and Outlook} 

We have undertaken a detailed study of the thresholds where multi-particle Kaluza-Klein gravitons
have order one correlations at large $N$ with single gravitons. 
The angular momenta of the gravitons  in $AdS_5 \times S^5 $ must grow large with $N$ for the correlator to approach the threshold, 
and the precise form of this growth was worked out in several cases.
The  large $N$ growth at the threshold region for the case of two gravitons of angular momentum $J$ being correlated with a single graviton of angular momentum $2J$ is  $J \approx \sqrt { N \log N }$. 
The breakdown of factorization is a breakdown of the usual perturbative scheme for computing graviton interactions 
in spacetime, which relies on a multi-graviton Fock space with states of different particle number being orthogonal. 
In this usual framework, the mixing between different particle numbers arises in $1/N$ corrections 
which are suppressed at large $N$ for small enough $J$.
We have found quantitative description of several factors  which can move a correlator from the regime
factorization  to the threshold, such as:
\begin{itemize} 
\item Increasing the total energy of the gravitons, 
\item Decreasing the separation in the energies of the two gravitons,
\item Decreasing the separation of gravitons in the boundary directions,
\item Increasing the number of gravitons.
\end{itemize} 
Another qualitative outcome of interest is that for $k$ gravitons being correlated with a single graviton, the
threshold can be   expressed in terms of the two largest 
momenta among the $k$ gravitons, taking the 
 form $ J_i J_j \approx N \log N $.  In these investigations, we have found  a rich variety of applications of 
the Lambert $W$-function. We have seen intriguing similarities between asymptotic threshold equations and running gauge 
couplings in non-abelian gauge theories. The large $N$ approximations have also involved transseries of the kind 
seen in instanton-corrected perturbation expansions of quantum field  theory. 

We also investigated the factorization thresholds in the presence of LLM backgrounds 
associated with rectangular Young diagram backgrounds. The presence of 
these backgrounds \emph{increases} the region of graviton momenta that are consistent 
with factorization. There are indications that triangular Young diagrams can be used to 
model thermal black hole-like backgrounds \cite{BBJS05}. We expect that, in the presence of 
black holes, the regime of validity of effective field theory should be smaller than in the absence of 
black holes. This would suggest that factorization in triangular Young diagram backgrounds 
should occur in a more limited regime of graviton angular momenta than factorization 
in the vacuum. This is a very concrete problem in the combinatorics of CFT correlators, 
and an interesting  research direction  for the future. 

In our study of factorization thresholds, we have consistently found thresholds when the 
angular momenta are of the form  $J_i J_j  \approx N \log N$, which suggests that there is some
form of \emph{universality} of the threshold. An interesting future direction would be to
consider the thresholds calculated in the `overlap-of-states' norm from \cite{bbns,BdMKRT}, 
as opposed to the `multiparticle' norm used in this paper. In the overlap normalization, the correlators 
are bounded by one from above and cannot grow exponentially with $N$, but they may well 
tend to a finite non-zero constant at large $N$ if their angular momenta grow quickly enough. 
We could define a threshold in the overlap normalization as the surface where a correlator 
is equal to some fixed constant between zero and one. Evidence from shifting the factorization
threshold at the end of Section \ref{sec:j1eqj2soln} suggest that the form of the threshold will not
change when going to the overlap norm, and will remain $J^2\approx N \log N$.
This is another interesting problem for the future that involves non-trivial asymptotics of finite
$N$ CFT correlators, and could well provide further evidence for the universality of the $N \log N$ threshold.

In Section \ref{sec:next} we showed how the `nearly-extremal' correlator has a threshold which depends 
on the separation of the CFT-insertions in the 4D spacetime directions, as well 
as exhibiting the dependences on total energy and energy differences  of the corresponding gravitons. 
We considered two gravitons in AdS with angular momenta $(J_3, J_1), (-J_3, J_2)$ 
where the  first entry refers to the $Y$-plane and the second to the $Z$-plane. We studied the 
correlation with a single graviton with angular momenta $ ( J_1 + J_2 , 0 )$. 
The explicit calculations were done for $J_3=1$, with $J_1 , J_2 $ growing  with $N$. 
A generalization to the case of $ J_3 $ also growing  with $N$ would be very interesting, 
as it would show the effect on the quantum correlations at threshold between two gravitons 
and a single graviton, when the two gravitons annihilate a large amount of $Y$-momentum 
and the correlator is no longer near-extremal.

We hope to have convinced the reader that the theme of thresholds between different behaviours is a fruitful way to 
explore the bulk AdS physics encoded in the correlators of the CFT. Since
\bea 
\frac{1}{N} = g_s \frac{l_s^4}{R^4 }  \, , 
\eea
for fixed $ { R /  l_s } $,  finite $N$ is  finite string coupling, which is 
non-perturbative from the point of view of strings in the bulk spacetime.
Hence, finite $N$  calculations in CFT contain valuable information about  strongly 
quantum gravitational effects.  The generic $ J_i J_j \approx N \log N$ we found, which in spacetime variables is 
\bea
J_i J_j  \approx N \log N = \left( \frac{  R^4 }{ g_s l_s^4}\right) \log \left( \frac{R^4 }{g_s l_s^4}\right),  \nn
\eea 
 is an intriguing result that should be understood better from the bulk point of view, 
either from a first principles string calculation in $AdS_5 \times S^5$  or  from a phenomenological model of quantum 
gravitational  spacetime constructed to reproduce  the CFT result.  As we observed, the threshold corresponds to a region where 
the Fock space of spacetime field  modes breaks down. The broader issue of the breakdown of perturbative 
effective field theory is central to questions in black hole physics \cite{PapRaju,susskthor,AMPS}.
In particular, black hole complementarity is related to the structure of Hilbert spaces needed to 
describe infalling observers and outgoing radiation. 
We propose that a convincing spacetime understanding of the thresholds derived here
would be a highly instructive step in understanding the departures from effective field theory in quantum gravity. 
Insights from earlier work on bulk spacetime  in AdS in connection with gauge-string duality,
such as in \cite{BGL99,HKLL06}, might be useful.  Alternatively, the methods of collective field theory \cite{DasJev} could help with   
a derivation of the large $N$ effective field theory. Another possible approach towards better understanding the thresholds
from the spacetime point of view would be to make use of  a combination of  semi-classical tools, 
exploiting high energy  eikonal approximations or  physical effects such as the tidal stretching of high energy gravitons into strings, for example along the lines of 
 \cite{DDRV,ASVW}.

The study of Schur operators as the description of giant gravitons was motivated by the observed
departure from orthogonality between multi-graviton and single graviton states at large $J$ \cite{bbns}.
Schur operators give a weakly-coupled description of giant gravitons in the regime of $J\approx N$, 
but become strongly-interacting as $J$ is decreased \cite{dms}.
In this paper, we have focused on the approach to the threshold in the regime 
near $J \approx \sqrt { N }$ by studying single and multi-trace graviton operators. 
It would be very interesting to study thresholds between weak and strong interactions in giant graviton physics as the 
angular momenta are decreased from $J \approx N$. The detailed investigations of the one-loop and multi-loop 
dilatation operators around giant graviton backgrounds should provide useful data for this purpose \cite{gigravosc,doubcos,KGM}.

The fact that the thresholds are near $\sqrt { N }$ rather than $N^{1/4}$ 
is rather intriguing. This has been discussed in \cite{dms}.  Angular momenta of $ J \approx  N^{ 1 / 4 } $ 
correspond to momenta comparable to the ten-dimensional Planck scale. 
This may be a sign that $AdS_5 \times S^5$ physics is just very different from expectations derived from 
effective field theory in flat space  $\mR^{ 9, 1 }$. On the other hand, it could be that
a clever interpretation of the link between the extremal correlators and flat space scattering 
would account for the thresholds we see from the CFT. 
Potentially, the correct interpretation has to recognise that extremal correlators
correspond to collinear graviton scatterings.
We would need to consider the flat space expectations in the light of collinear effective theories 
of gravitons, along the lines developed in \cite{ASS}, to understand the difference between
the threshold scale and the Planck scale.
An early discussion of the subtleties of connecting bulk AdS spacetime  physics to  the flat space limit is given in \cite{Gidd99}. 

There is a lot of fun to be had with factorization thresholds in AdS/CFT: there is a wealth of 
quantitative information about graviton correlations at threshold  available via finite $N$  CFT computations
and their large $N$ asymptotics. The lessons we draw from these are very likely to be important for questions we would like to answer in black hole physics and quantum gravity.


\begin{center}
{\bf Acknowledgements}
\end{center}
It is a pleasure to thank Pawel Caputa, Robert de Mello Koch, Yang Hui He,  Robert Myers, Gabriele Travaglini, and Donovan Young
for discussions. This work was supported by the Science and Technology Facilities
Council Consolidated Grant ST/J000469/1 String Theory, Gauge Theory and Duality.

\section*{Appendix}
\appendix
\section{The Lambert \texorpdfstring{$W$}{W}-function}

The Lambert $W$-function is, by definition, the solution to the equation
\bea
W(z)e^{W(z)} = z. \label{eq:wdef}
\eea
This equation cannot be solved in a closed form in terms of elementary functions, but 
a Taylor series can be found near $z=0$, and its asymptotic series can be derived for large
positive $z$.

\begin{figure}[t]
\centering
\includegraphics[width=0.5\textwidth]{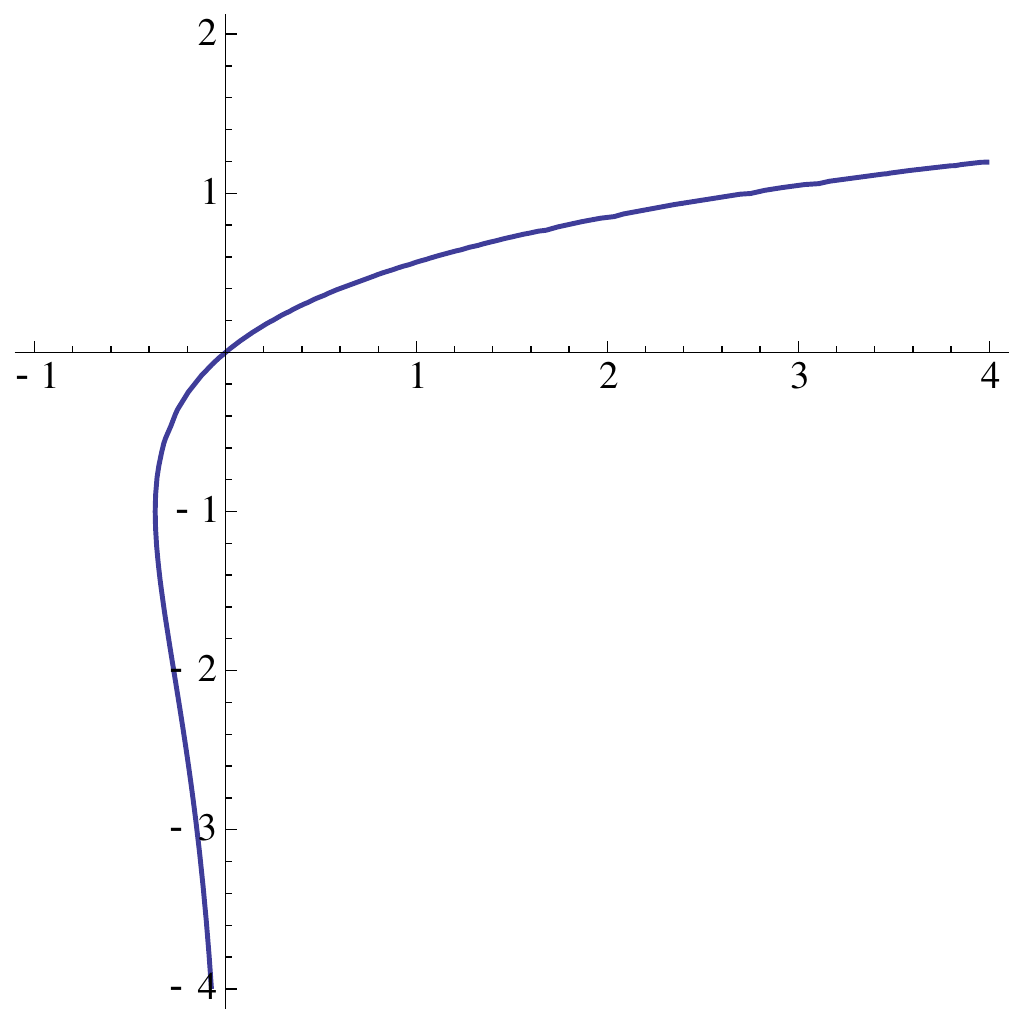}
\caption{The Lambert $W$-function $W(x)$ for real $x$ is multivalued: the principal branch $W_0$ takes values greater than -1,
and the other branch $W_{-1}$ is defined for $W<-1$.}
\label{fig:lambertw}
\end{figure}

There are many solutions $W(z)$ to the equation \refb{eq:wdef}, which means that
the Lambert $W$-function is multivalued. However, only two solutions
take real values when $z$ is real, and these are the only relevant solutions in this paper.
One of these solutions is the principal branch $W_0(z)$, which is real 
and satisfies $W_0(z) \geq -1$ on its domain $z\in [-e^{-1}, \infty)$.
The other is the $W_{-1}(z)$ branch, which takes values in the range
$W_{-1}(z)\leq -1$ and is defined on the domain $z \in [-e^{-1}, 0)$.
The two real branches of the $W$-function are shown in figure \ref{fig:lambertw}.

The large $z$ expansion of the principal branch of the $W$-function is 
\bea
W(z) \sim \log z - \log \log z +\sum_{n=1}^\infty \left(\frac{-1}{\log z}\right)^n \sum_{k=1}^n \stirling{n}{n - k + 1} \frac{(-\log \log z)^k}{k!}, \label{eq:wseries2}
\eea
where the coefficients in the square brackets are the (unsigned) Stirling cycle numbers of the first kind.
The notation $\stirling{n}{k}$ denotes the number of permutations of $n$ elements composed of $k$ disjoint cycles.
(For example, $\stirling{4}{2}$ refers to the number of permutations in the symmetric group $S_4$ composed of two disjoint cycles.
There are six permutations in $S_4$ composed of a 3-cycle and a 1-cycle, and three permutations composed of a pair of disjoint
2-cycles, and these are the only permutations composed of two disjoint cycles in $S_4$. Hence, $\stirling{4}{2}=6+3=9$.)

\section{Combinatoric calculations using character sums}\label{sec:appmm}

In this appendix we present some finite $N$ calculations of correlators using matrix model techniques.
The extremal correlator $\cor{\tr Z^{J_1} \tr Z^{J_2} \tr Z^{\dagger J_1+J_2}}$ was calculated 
in \cite{kpss}, and using character sums in \cite{cr}. We use the methods of \cite{cr} to calculate 
the norm of the operator $ \str(Z^{J_1} Y^{J_2})$, and to calculate the $k\to 1$ correlator
$\cor{\tr Z^{J_1}\tr Z^{J_2} \ldots \tr Z^{J_k} \tr Z^{\dagger \sum_i J_i}}$. We then find an 
expression for the normalized $k+1$-point correlator at large $N$.

\subsection{The non-extremal operator norm}

Consider the non-extremal two-point function which is the norm of a mixed operator consisting of two types of adjoint fields,
\bea \norm{ \str(Z^{J_1} Y^{J_2}) }^2 &=& \cor{\str(Z^{J_1} Y^{J_2}) \str(Z^{\dagger J_1} Y^{\dagger J_2}) }.\eea
The symmetrized trace of a string of matrices in the adjoint representation of the gauge group $U(N)$ is
\bea
\str(Z^{J_1}Y^{J_3}) = \frac{1}{(J_1+J_3-1)!}\sum_{\sigma \in [J_1+J_3]} X^{i_1}_{i_{\sigma(1)}}
X^{i_2}_{i_{\sigma(2)}} \ldots
X^{i_{J_1}}_{i_{\sigma(J_1)}}
Y^{i_{J_1+1}}_{i_{\sigma(J_1+1)}}\ldots
Y^{i_{J_1+J_3}}_{i_{\sigma(J_1+J_3)}}.
\eea
The sum is performed over all permutations in $[J_1+J_3]$, the conjugacy class in $S_{J_1+J_3}$ consisting of all 
the cyclic permutations with a single cycle of length $(J_1+J_3)$.
All matching pairs of adjoint matrix indices $i_l$ are implicitly summed.
This expression can be written more concisely in tensor space notation \cite{cr} as
\bea
\str(Z^{J_1}Y^{J_3}) = \frac{1}{(J_1+J_3-1)!}\sum_{\sigma \in [J_1+J_3]} \tr (\sigma X^{\otimes J_1}\otimes Y^{\otimes J_3}).
\eea

\begin{figure}[t]
\centering
\includegraphics[width=0.35\textwidth]{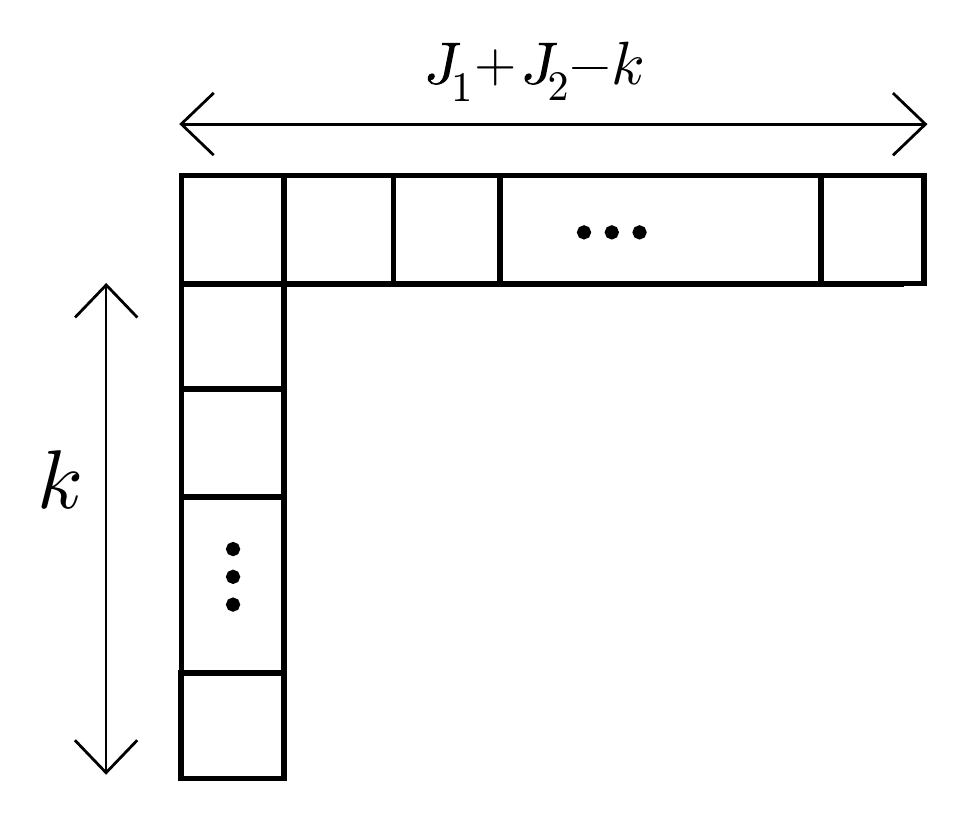}
\caption{A Young diagram with $J_1+J_2$ boxes corresponding to a hook rep with hook length $k$.}
\label{fig:hook}
\end{figure}

This two-point function can be calculated by using diagrammatic tensor space techniques \cite{cr}:
\bea
\norm{ \str(Z^{J_1} Y^{J_2}) }^2 &=& \frac{1}{(J_1+J_2-1)!^2}\sum_{\sigma_1, \sigma_2 \in [J_1+J_2] } \raisebox{-50pt}{\includegraphics[height=100pt]{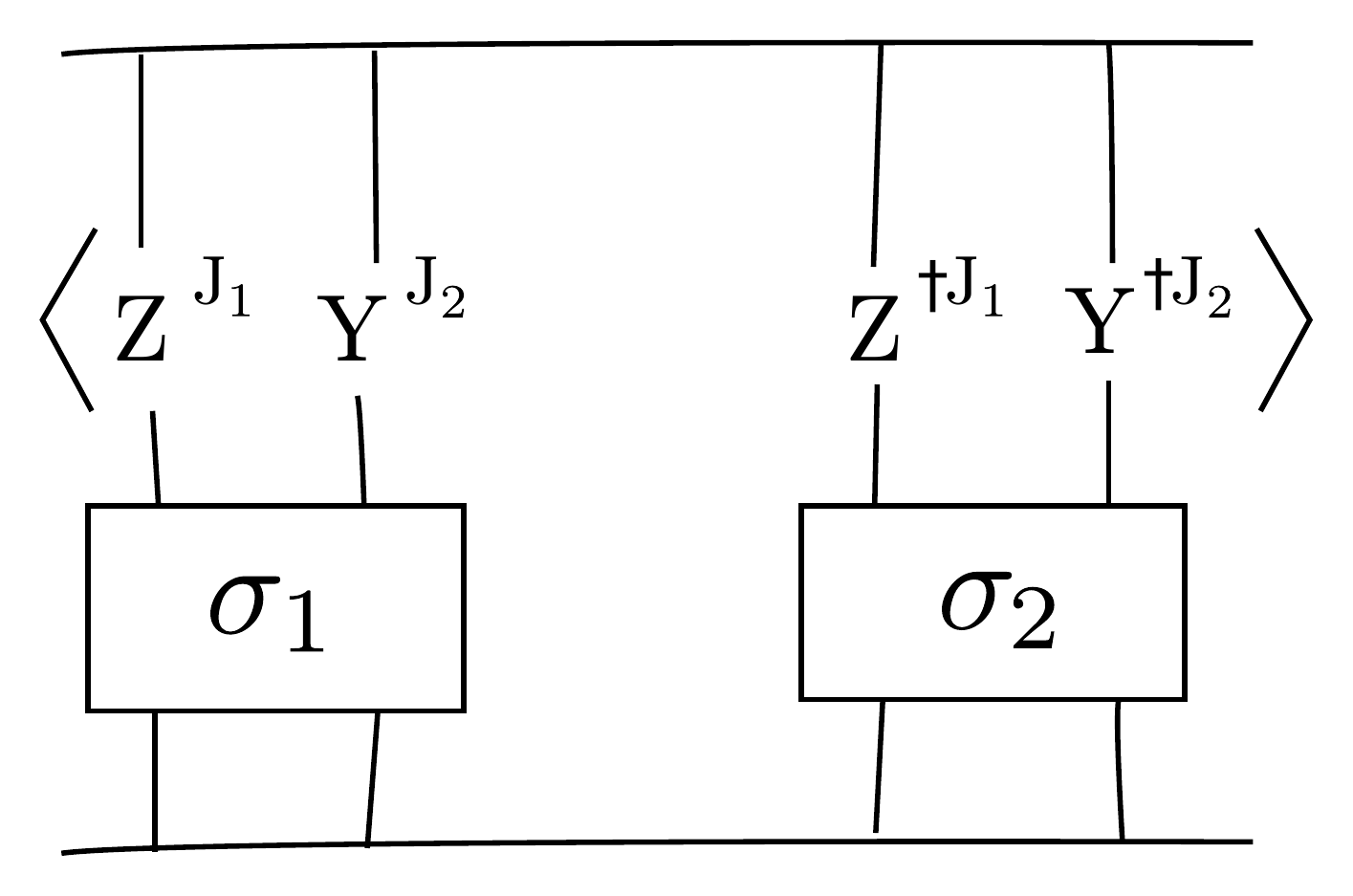}} \label{ebox1}
\eea
\bea
= \frac{1}{(J_1+J_2-1)!^2}\sum_{\substack{\sigma_1, \sigma_2 \in [J_1+J_2] \\ \gamma_1\in S_{J_1}\\ \gamma_2\in S_{J_2}} }\raisebox{-50pt}{\includegraphics[height=100pt]{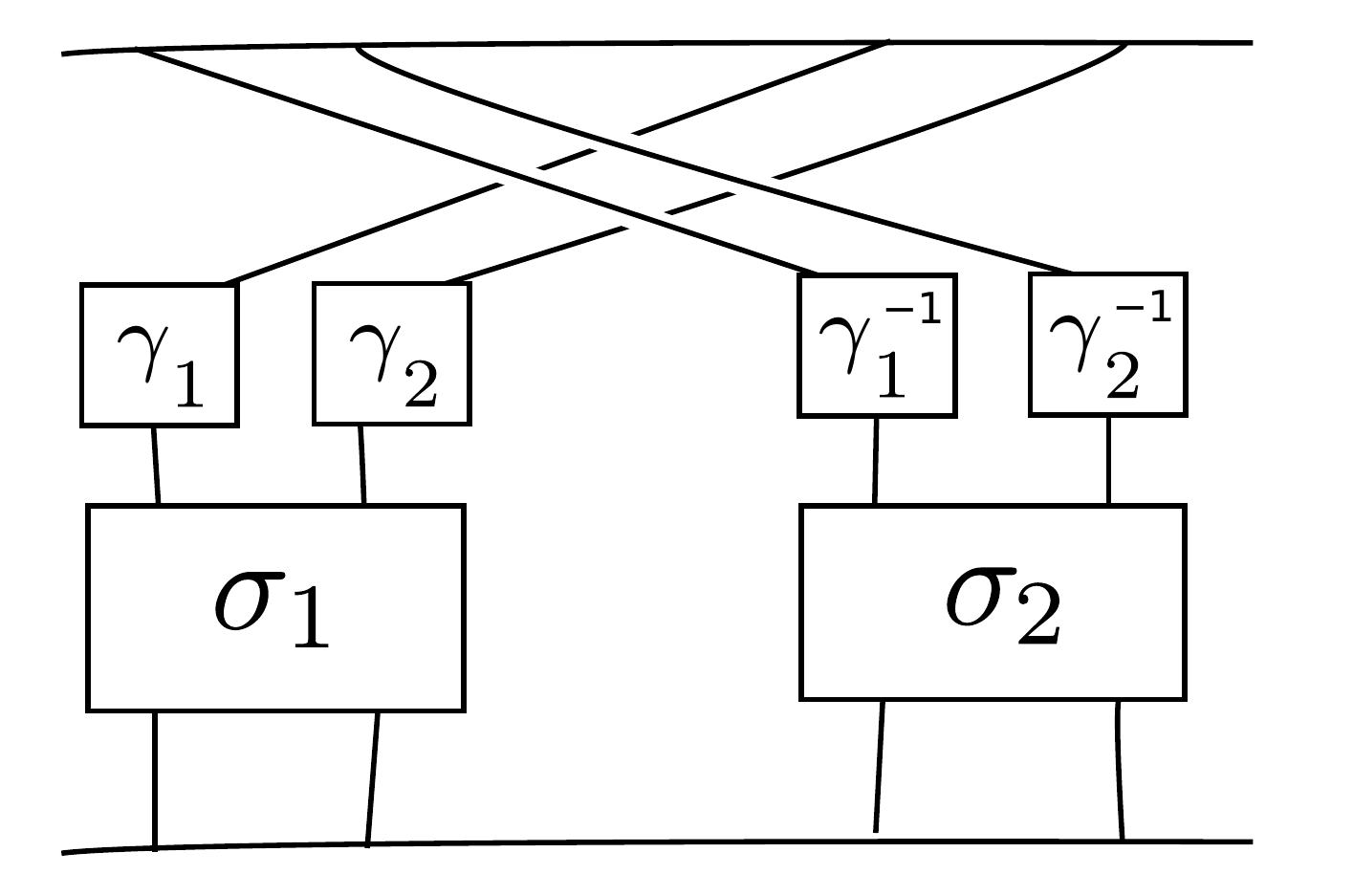}} \eea
\bea
\frac{J_1!J_2!}{(J_1+J_2-1)!^2}\sum_{\sigma_1, \sigma_2 \in [J_1+J_2]}\raisebox{-50pt}{\includegraphics[height=100pt]{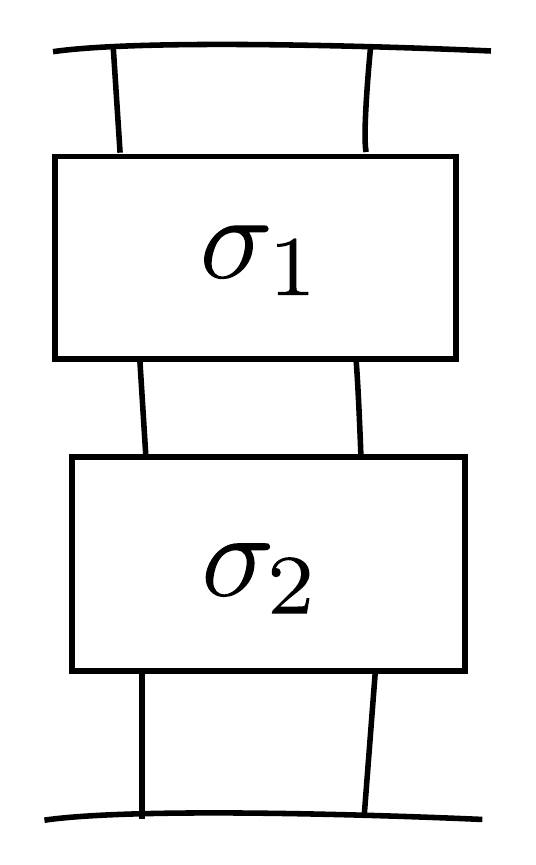}}
\eea
We can replace the permutation sums with sums over representations with projectors on the group algebra,
\bea
\norm{ \str(Z^{J_1} Y^{J_2}) }^2 = J_1!J_2!\sum_{\substack{R_1, R_2 \vdash (J_1+J_2) }} \frac{\chi_{R_1}([J_1+J_2])\chi_{R_2}([J_1+J_2])}{d_{R_1}d_{R_2}}\raisebox{-50pt}{\includegraphics[height=100pt]{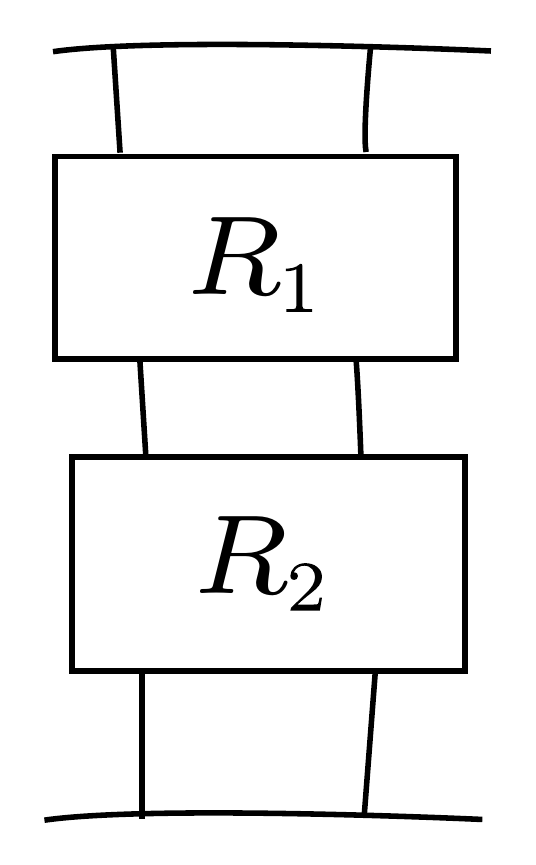}},
\eea
where $\chi_{R_1}([J_1+J_2])$ is the character in $R_1$ of a permutation in the conjugacy class $[J_1+J_2]$.
Representation projectors satisfy the identity $P_{R_1}P_{R_2}=\delta_{R_1R_2}P_{R_1}$, and $\tr P_R = dim_N(R) d_R$. 
From the Murnaghan-Nakayama lemma \cite{fultonharris}, the character of a $(J_1+J_2)$-cycle in $S_{J_1+J_2}$  is $\pm1$ if the diagram is a \emph{hook}, and zero otherwise.
A hook representation corresponds to a Young tableau where all the boxes are in the first row or the first column, as in
Figure \ref{fig:hook}. 
We find
\bea
\norm{ \str(Z^{J_1} Y^{J_2}) }^2 &=& J_1!J_2!\sum_{R\vdash (J_1+J_2)}\frac{\chi_R([J_1+J_2])^2}{d_R^2}\tr(P_R) \\
&=& J_1!J_2! \sum_{R\text{\ a hook rep}} \frac{dim_N(R)}{d_R}. 
\eea
This sum is weighted by the dimension of a hook rep of $U(N)$ divided by the dimension of the corresponding
hook rep in $S_{J_1+J_2}$. Parametrizing the hook lengths by the hook length $k$, where $k=0,1,\ldots (J_1+J_2-1)$,
we find that the ratio of the dimensions is 
\bea 
\frac{dim_N(R)}{d_R}  = \binom{N+J_1+J_2-k -1}{J_1+J_2}, \eea
and hence the correlator is
\bea
\norm{ \str(Z^{J_1} Y^{J_2}) }^2 &=& J_1!J_2! \sum_{k=0}^{J_1+J_2-1}\binom{N+J_1+J_2 - k - 1}{J_1+J_2} \\
&=&  J_1!J_2!\sum_{k=0}^{J_1+J_2-1}\binom{N+k}{J_1+J_2}. \eea
Finally, we employ the general identity
\bea \sum_{k=0}^{n-1}\binom{N+k}{m} = \binom{N+n}{m+1} - \binom{N}{m+1} \label{binid1} \eea
to deduce the final exact answer,
\bea
\norm{ \str(Z^{J_1} Y^{J_2}) }^2 \  = J_1!J_2! \left[ \binom{N+J_1+J_2}{J_1+J_2+1} - \binom{N}{J_1+J_2+1} \right]. \label{final1}
\eea

\subsection{The \texorpdfstring{$k+1$}{k+1}-graviton correlator character sum}

In this section we present a calculation of the $k+1$-graviton correlator in the gauge theory. A similar
calculation was done previously in \cite{CaputaMohammed}.
The representation sum of the general extremal correlator was stated in \cite{cr} as being
\bea
\cor{ \prod_{i=1}^k (\tr Z^{J_i}) \tr Z^{\dagger J}} = \sum_{R\vdash J} f_R \chi_R([J_1\ldots J_k])\chi_R([J]).
\eea
We adopt the notation $J= \sum_i J_i$ throughout this subsection.
Using the Murnaghan-Nakayama lemma \cite{fultonharris}, we find that $\chi_R([J])$ is non-zero only 
if $R$ is a hook rep, and equal to $(-)^k$ for a hook of length $k$. This constrains the sum to run only over
hook representations, and so
\bea
\cor{ \prod_{i=1}^k (\tr Z^{J_i}) \tr Z^{\dagger J}} &=& J!\sum_{l=0}^{J-1} \binom{N+l}{J}(-)^{J-1-l}\chi_{H_l}([J_1\ldots J_k]), \label{recursive}
\eea
where $H_l$ denotes the hook representation $[l+1, 1^{J-1-l}]$.
The Murnaghan-Nakayama lemma states that we can knock $J_k$ boxes off this $J$-box hook rep to get
\bea
\chi_{H_l}[J_1\ldots J_k] = \delta(l\geq J_k)\chi_{H_{l-J_k}}([J_1\ldots J_{k-1}]) + (-)^{J_k+1}\delta(J-l>J_k)\chi_{H_l}([J_1\ldots J_{k-1}])
\eea
If we replace the expressions in the binomial coefficient by the general terms $M$, $m$, we have
\begin{multline*}
\sum_{l=0}^{J-1} (-)^l \binom{M+l}{m}\chi_{H_l}([J_1\ldots J_k]) = \sum_{l=0}^{J-J_k-1}(-)^{J_k}\chi_{H_l}([J_1\ldots J_{k-1}]) \left[ \binom{M+J_k+l}{m} - \binom{M+l}{m}\right]
\end{multline*}
We can plug this formula in to \refb{recursive} for different values of $M$ and $m$. We get
\bea
\cor{ \prod_{i=1}^k (\tr Z^{J_i}) \tr Z^{\dagger J}} = J!(-)^{J-1}\sum_{l=0}^{J-1}(-)^{l}\chi_{H_l}([J_1\ldots J_k]) \binom{N+l}{J}
\ret
= J!(-)^{J-J_k-1}\sum_{l=0}^{J-J_k-1}(-)^l\chi_{H_l}([J_1\ldots J_{k-1}])\left[\binom{N+J_k+l}{J} - \binom{N + l}{J}\right] \nn
\eea
\begin{multline*}
= J!(-)^{J-J_k-J_{k-1}-1}\sum_{l=0}^{J-J_k-J_{k-1}-1}(-)^l\chi_{H_l}([J_1\ldots J_{k-2}])\left[\binom{N+J_k+J_{k-1}+l}{J} 
- \binom{N+J_k+l}{J} \right. \\ \left. - \binom{N+J_{k-1}+l}{J} + \binom{N + l}{J}\right] 
\end{multline*}
\bea
= J!(-)^{J_1-1}\sum_{l=0}^{J_1-1} (-)^l\chi_{H_l}([J_1]) \left[\binom{N + J-J_1 + l}{J} - \ldots + (-)^{k-1}\binom{N+l}{J}\right],
\eea
where we have omitted the intermediate binomials with arguments containing all sums of elements in $\{J_2, J_3 \ldots, J_k\}$. Using $\chi_{H_l}([J_1]) = (-)^{J_1-1-l}$ and 
\bea
\sum_{l=0}^{J_1-1}\binom{M+l}{J} = \binom{M+J_1}{J+1} - \binom{M}{J+1},
\eea
we can now evaluate the sums to find that
\bea 
\cor{ \prod_{i=1}^k (\tr Z^{J_i}) \tr Z^{\dagger J}} = J!\left[ \binom{N + J}{J+1} -\ldots + (-)^k \binom{N}{J+1}\right]
\eea
and restoring the omitted terms, we deduce that
\bea
\cor{ \prod_{i=1}^k (\tr Z^{J_i}) \tr Z^{\dagger J}} = J!\sum_{t=0}^k \sum_{\substack{S\subseteq\{1,\ldots, k\}\\ |S|=t}}(-)^{k-t}\binom{N+ \sum_{i\in S}J_i}{J+1}. \label{eq:kpoint}
\eea
The sum over $S$ is a sum over all the subsets of the $k$-element set.

\subsection{Asymptotics of the \texorpdfstring{$k+1$}{k+1}-point function}
In this section we derive the asymptotic form of the $k+1$-point function \refb{eq:kpoint}.
Assuming that $J_i\ll N^\frac{2}{3}$ and that $\Lambda$ is some sum of the $J_i$, we have from Section \ref{sec:j1eqj2}
\bea
J!\binom{N+\Lambda}{J+1} 
&\sim & \frac{N^{J+1}}{J}\exp\left(\frac{J(2\Lambda-J)}{2N} -  \cO\left(\frac{J^3}{N^2}\right) \right).
\eea
We can then write \refb{eq:kpoint} as
\bea
\cor{\prod_{i=1}^k (\tr Z^{J_i}) \tr Z^{\dagger J}} &\sim & \frac{N^{J+1}}{J}\sum_{S \subseteq \{1\ldots k\}}(-)^{k-|S|}e^{-\frac{J^2}{2N} + \cO(J^3/N^2)}
e^{\frac{J}{N}\sum_{i\in S}J_i} \\
&\sim & \frac{N^{J+1}}{J}e^{-\frac{J^2}{2N}+  \cO(J^3/N^2)}(-)^k\sum_{S\subseteq \{1\ldots k\}} \prod_{i\in S}\left(-e^{\frac{J}{N}J_i}\right)
\eea
We can evaluate this sum over subsets explicitly by first partitioning the sum into two; one sum over the subsets including the element
$k$, and one over the subsets not including $k$. We can then apply this for each integer from $1$ to $k$. We have
\bea
\sum_{S \subseteq \{1\ldots k \}}\prod_{i\in S}\left(-e^{\frac{J}{N}J_i}\right) &=& \left(-e^{\frac{J}{N}J_k}\right) \sum_{S \subseteq \{1\ldots k-1 \}}\prod_{i\in S}\left(-e^{\frac{J}{N}J_i}\right) + 1\sum_{S \subseteq \{1\ldots k-1 \}}\prod_{i\in S}\left(-e^{\frac{J}{N}J_i}\right) \\
&=& (-e^{\frac{J}{N}J_1} + 1)(-e^{\frac{J}{N}J_2} + 1)\ldots(-e^{\frac{J}{N}J_k} + 1).
\eea
Taking out a factor of $e^{J^2/N}$ from this product, we have the asymptotic form of the unnormalized correlator,
\bea
\cor{\prod_{i=1}^k (\tr Z^{J_i}) \tr Z^{\dagger J}} \sim \frac{N^{J+1}}{J}\exp\left(\frac{J^2}{2N} + \hot{J^3}{N^2}\right)\prod_{i=1}^k\left(1- e^{-\frac{JJ_i}{N}}\right).
\eea
Together with the known asymptotic form of the 2-point function
\bea
\cor{\tr Z^{J_i} \tr Z^{\dagger J_i}} &\sim& \frac{N^{J_i+1}}{J_i}e^{\frac{J_i^2}{2N}}\left(1-e^{-\frac{J_i^2}{N}}\right),
\eea
we can therefore write the full correlator in the large $J$, small $J^3/N^2$ limit,
\bea
\corb{\prod_{i=1}^k (\tr Z^{J_i}) \tr Z^{\dagger J}} \sim \sqrt{\frac{J_1\ldots J_k}{JN^{k-1}}}e^{\frac{J^2}{4N}-\frac{J_1^2}{4N}-\ldots - \frac{J_k^2}{4N}} \frac{\prod_{i=1}^k(1-e^{-JJ_1/N})}{\sqrt{(1-e^{-J^2/N})\prod_{i=1}^k(1-e^{-J_1^2/N})}} \\
\sim \sqrt{\frac{J_1\ldots J_k}{JN^{k-1}}} \frac{(1-e^{-\frac{JJ_1}{N}})\ldots (1-e^{-\frac{JJ_k}{N}})}{ \sqrt{(1-e^{-\frac{J_1^2}{N}})\ldots (1-e^{-\frac{J_k^2}{N}})(1-e^{-\frac{J^2}{N}})} } \exp\left( \sum_{i<j} \frac{J_i J_j}{2N} + \hot{J^3}{N^2}\right).
\eea
When all the angular momenta are equal, $J_1=J_2=\ldots=J_k = J$, and $J^2/N$ is large, then this expression becomes
\bea
\corb{ (\tr Z^{J})^k \tr Z^{\dagger kJ}} \sim \sqrt{\frac{J^{k-1}}{kN^{k-1}}}\exp\left(\frac{k(k-1)}{4}\frac{J^2}{N}\right).
\eea

\end{document}